\documentclass[trackchanges]{aastex701} 

\accepted{}

\submitjournal{ApJ}

\usepackage{float}
\usepackage{comment}
\usepackage{upgreek}
\usepackage[T1]{fontenc}
\usepackage{hyperref}
\usepackage{subcaption}
\usepackage{soul}
\usepackage{threeparttable}

\begin{document}

\title{Exploring Polarized Millimeter Emission from Protoplanetary Disks with Irregular Dust Grains}

\correspondingauthor{Jes\'us Miguel J\'aquez-Dom\'inguez}
\author[orcid=0000-0002-1912-5394]{Jes\'us Miguel J\'aquez-Dom\'inguez}
\affiliation{Instituto de Radioastronom\'ia y Astrof\'isica, Universidad Nacional Aut\'onoma de M\'exico, Antigua Carretera a P\'atzcuaro \# 8701, Ex-Hda. San Jos\'e de la Huerta, Morelia, Michoac\'an, M\'exico C.P. 58089}
\email[show]{j.jaquez@irya.unam.mx}

\author[orcid=0000-0003-2862-5363]{Carlos Carrasco-González}
\affiliation{Instituto de Radioastronom\'ia y Astrof\'isica, Universidad Nacional Aut\'onoma de M\'exico, Antigua Carretera a P\'atzcuaro \# 8701, Ex-Hda. San Jos\'e de la Huerta, Morelia, Michoac\'an, M\'exico C.P. 58089}
\email[]{c.carrasco@irya.unam.mx}

\author[orcid=0000-0002-9228-1035]{Daniel Guirado}
\affiliation{Instituto de Astrof\'isica de Andaluc\'ia (IAA-CSIC), Glorieta de la Astronom\'ia s/n, E-18008 Granada, Spain}
\email[]{dani@iaa.es}

\author[orcid=0000-0002-5138-3932]{Olga Muñoz}
\affiliation{Instituto de Astrof\'isica de Andaluc\'ia (IAA-CSIC), Glorieta de la Astronom\'ia s/n, E-18008 Granada, Spain}
\email[]{olga@iaa.es}

\author[orcid=0000-0003-1283-6262]{Enrique Macías}
\affiliation{European Southern Observatory, Karl-Schwarzschild-Str. 2, 85478 Garching bei München, Germany}
\email[]{enrique.macias@eso.org}

\author[orcid=0009-0000-5520-4455]{Gonzalo Vargas}
\affiliation{Instituto de Astrof\'isica de Andaluc\'ia (IAA-CSIC), Glorieta de la Astronom\'ia s/n, E-18008 Granada, Spain}
\email[]{gvargas@iaa.es}

\author[orcid=0000-0003-2211-4001]{Julia Martikainen}
\affiliation{Instituto de Astrofísica de Canarias, E-38200 La Laguna, Tenerife, Spain}
\affiliation{Nordic Optical Telescope, Rambla José Ana Fernández Pérez 7, E-38711 Breña Baja, Spain}

\email[]{juliamar@iac.es}

\begin{abstract}
Polarization at millimeter wavelengths provides a powerful diagnostic of dust grain properties in protoplanetary disks.
Standard  models based on solid spherical grains often struggle to reproduce the observed polarization fractions and morphologies in systems where self-scattering is expected to dominate. 
We investigate the impact of grain morphology on polarized millimeter emission by comparing models that adopt solid spherical grains with models that employ solid irregular hexahedral particles drawn from the \texttt{TAMUdust2020} database. 
Both grain populations share identical size distributions, enabling us to isolate the effects of geometry while preserving the same internal structure and material density. 
We explore three optical-depth regimes—optically thick, optically thin, and an intermediate hybrid case—to assess how grain morphology modifies the polarization structure under different conditions. For size distributions with $a_{\mathrm{max}} \sim \lambda / 2\pi$, where scattering-induced polarization is expected to peak, we find that the polarization morphology and fraction are nearly indistinguishable between spherical and irregular grains.
The primary quantitative difference is an enhancement of the scattering opacity by up to a factor of $\sim 2.5$ for irregular particles, implying that disk dust masses inferred under the assumption of spherical grains may be systematically overestimated.
Irregular grains also suppress the polarization reversal predicted by Mie theory at large size parameters ($x>1$).
Nevertheless, modifying grain geometry alone is insufficient to reproduce the observed polarization fractions within a pure self-scattering framework. These results suggest that additional physical effects, such as dust porosity, warrant dedicated investigation.
\end{abstract}


\section{Introduction}\label{sec:introduction}

Planetary systems emerge from the evolution of protoplanetary disks around young stars \citep[e.g.,][]{Keppler2018,Muller2018,Haffert2019}. During this evolution, submicron-sized interstellar grains grow, fragment, and migrate within disks whose temperature, density, and turbulence regulate dust dynamics. Although substantial theoretical progress has been achieved, the pathway from micron-sized particles to kilometer-scale planetesimals remains debated \citep[see][for a review]{Drazkowska2023ASPC}. The advent of high-sensitivity  ($\sim 10~\mathrm{\mu Jy \ beam^{-1}}$) and high-angular-resolution ($\sim0\farcs01$) observations with the Atacama Large Millimeter/submillimeter Array (ALMA) has revolutionized the field, resolving ring–gap substructures, spirals, and compact inner disks down to a few astronomical units in systems as young as $\sim1$ Myr \citep{ALMA_Brogan2015,Andrews2018DSHARP_I,Long2018}. These observations provide critical constraints on dust mass, radial segregation, and grain growth \citep{Carrasco-Gonzalez2016,Carrasco-Gonzalez2019ApJ...883...71C,Sierra2020ApJ...892..136S,Liu2024}.

Dust grain sizes are commonly inferred through spectral energy distribution (SED) modeling and through the wavelength dependence of polarized emission. Under the assumption of solid spherical grains, polarization modeling typically suggests maximum grain sizes of tens to a few hundred microns \citep{Kataoka2015,Kataoka2016a,Stephens2017}, whereas SED analyses require millimeter- to centimeter-sized particles \citep{Perez2012,Carrasco-Gonzalez2019ApJ...883...71C,Macias2021A&A...648A..33M}. This discrepancy indicates that the simplifying assumption of spherical grains may bias the interpretation of disk observations. At the same time, polarization measurements reveal a diversity of morphologies and wavelength-dependent behaviors \citep{Kataoka2016a,Stephens2017,Ohashi2018,Lin2023,Stephens2023}, suggesting that multiple mechanisms—including self-scattering \citep{Kataoka2015,Kataoka2016b}, magnetic alignment \citep{Cho2007,Hughes2009,yang2016b}, radiative alignment \citep{Tazaki2017}, and mechanical alignment \citep{Kataoka2019, Lin2024aligned} may coexist \citep{Yang2019}.

Despite this complexity, most radiative-transfer studies of disk polarization still rely on solid spherical grains. While spheroidal models introduce geometric anisotropy and permit intrinsic polarization \citep{Kirchschlager2019,Kirchschlager2020,ReyesAmador2024}, their optical behavior at millimeter wavelengths remains broadly similar to that of spheres and cannot reproduce the highest polarization fractions observed in several disks \citep{Kataoka2016b,Kataoka2017,Harrison2019ApJ...877L...2H,Lin2024}. Laboratory measurements and theoretical studies demonstrate that irregular and aggregate grains exhibit markedly different scattering phase functions and polarization efficiencies \citep{MUINONEN20091628,Escobar-Cerezo2017ApJ,Munoz2021ApJS..256...17M,Tazaki2016,Tazaki2019, Zhang2023, Potapov2025}. However, systematic radiative-transfer comparisons between spherical and truly irregular grains under varying optical-depth conditions remain limited \citep{Lin2023}.

These limitations motivate a dedicated investigation of how grain irregularity alone affects polarization produced by dust self-scattering. In this work, we deliberately exclude alignment mechanisms—whose efficiency for millimeter-sized grains in dense disk environments remains uncertain \citep{Kataoka2015,Kirchschlager2019}—in order to isolate the role of grain structure. Observations of interstellar and Solar System dust indicate that real particles are predominantly irregular rather than perfectly spherical \citep{MUINONEN20091628}, underscoring the need to evaluate the impact of more realistic morphologies on disk polarization.

To address this problem, we perform three-dimensional Monte Carlo radiative-transfer simulations using \texttt{RADMC-3D} \citep{Dullemond2012ascl.soft02015D}. We compare models that employ spherical grains with models that adopt irregular hexahedral particles whose optical properties are taken from the \texttt{TAMUdust2020} database  \citep{Saito2021JAtS...78.2089S}. By exploring optically thin, optically thick, and intermediate regimes, we quantify how grain morphology modifies the polarization fraction and spatial structure of the emergent emission. Our goal is to determine whether irregular grains alleviate the tension between polarization- and SED-derived grain sizes and to establish a more physically grounded framework for interpreting millimeter polarization in protoplanetary disks.

The structure of this paper is as follows. In Section \ref{sec:methods}, we describe how the optical properties of the dust models were obtained for both spherical and irregular grains. Additionally, the disk models used to test the polarized features in a realistic system are described. Section \ref{sec:results_discussion} presents the resulting polarization fraction maps and compares the polarized features between models using spherical and irregular grains in various optical depth regimes. Section \ref{sec:conclusions} provides a summary of the key findings.

\section{Methods}\label{sec:methods}

The aim of this study is to compare polarized emission from disk models populated with either solid spherical or solid irregular grains. The first step was to obtain the single-scattering properties of both grain populations, which are then used as inputs for radiative transfer simulations that account for multiple scattering in disks of different optical depths. For both spherical and irregular grains, we adopt the dust composition used in the Disk Substructures at High Angular Resolution Project (DSHARP; \citealt{Birnstiel2018}), consisting of 20\% water ice, 33\% astronomical silicates, 7\% troilite, and 40\% refractory organics by mass. The resulting bulk density of the mix is $\rho = 1.675 \,\mathrm{g\,cm^{-3}}$. We perform this study at a fixed wavelength of 1 mm, at which the corresponding refractive index is $m = 2.2995 + i\,0.02031$, calculated using \texttt{dsharp\_opac}\footnote{\url{https://github.com/birnstiel/dsharp_opac}} \citep{Birnstiel2018}.

For spherical particles, the optical properties, including absorption and scattering opacities and the full scattering matrix, were computed using the Mie theory implemented in the \textsc{Optool}\footnote{\url{https://github.com/cdominik/optool}} package \citep{Dominik2021ascl.soft04010D}. We assume a power-law size distribution with an index $q=-3.5$, extending from $a_\mathrm{min}=1\,\mu$m to $a_\mathrm{max}=160\,\mu$m. This size distribution maximizes the efficiency of self-scattering polarization at millimeter wavelengths, i.e., $a_\mathrm{max} \sim \lambda/2\pi$ for $\lambda \sim 1$ mm \citep{Kataoka2015, Kataoka2016a}. The resulting mass opacities are shown in Table \ref{tab:opacities}. 

\begin{table*}
    \centering
    \begin{threeparttable}
        \caption{Mass opacities of the different grain models used in this work.}
        \begin{tabular}{lcccc}
            \hline
            Grain shape & $a_\mathrm{max} \ (\mathrm{\mu m})$ & $k_\mathrm{ext} \ (\mathrm{cm^2\,g^{-1}})$  & $k_\mathrm{sca} \ (\mathrm{cm^2\,g^{-1}})$ &  $k_\mathrm{abs} \ (\mathrm{cm^2\,g^{-1}})$  \\
            \hline
            Spheres   & 160 & 5.41 & 4.57 & 0.84 \\
            Irregular\tnote{a} & 160 & 8.41 & 7.20 & 1.20 \\
            Irregular\tnote{b} & 160 & 13.90 & 12.52 & 1.38 \\
            \hline
            Spheres & 300 & 25.83 & 24.03 & 1.8 \\
            Irregular\tnote{a} & 300 & 38.65 & 36.38 & 2.27 \\
            \hline
            Spheres & 500 & 28.82 & 26.23 & 2.59 \\
            Irregular\tnote{a} &500 & 53.21 & 49.63 & 3.58 \\
            \hline
            Spheres & 1000 & 25.1 & 22.4 & 2.7 \\
            Irregular\tnote{a} & 1000 & 36.74 & 32.55 & 4.19 \\
            \hline
        \end{tabular}
        \begin{tablenotes}
            \item[a] Radius based on the projected-surface area. Used during this work.
            \item[b] Radius based on the equivalent-volume.
        \end{tablenotes}
        \label{tab:opacities}
    \end{threeparttable}
\end{table*}

\begin{figure}[ht]
    \centering
    \includegraphics[width=0.5\linewidth]{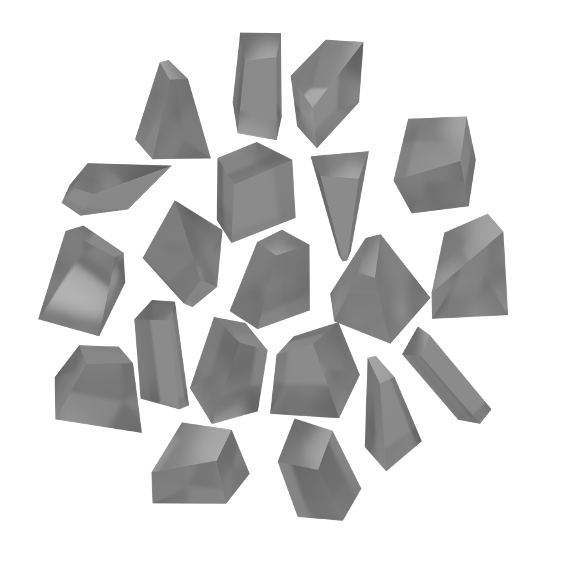}
    \caption{An example of the ensemble irregular hexahedra particles used by the TAMUdust2020 database to calculate the properties of the solid irregular particles used in our models. Figure adapted from TAMUdust2020 web page\footref{tamudustweb}.}
    \label{fig:irregular_particles}
\end{figure}

For the irregular particles, we employ an ensemble of hexahedral grains whose geometric and scattering properties are obtained from the \texttt{TAMUdust2020}\footnote{\label{tamudustweb}\url{https://sites.google.com/site/masanorisaitophd/dataresources/scattering-databases/tamudust2020}} database   \citep{Saito2021JAtS...78.2089S, masanori_saito_2021_4711247}. 
This database provides precomputed single-scattering optical properties (including full phase matrices) for ensembles of 20 irregular hexahedral particles (see Figure \ref{fig:irregular_particles}), spanning wide ranges of size parameter, complex refractive index, particle sphericity, and wavelength for use in radiative transfer simulations. The ability of the \texttt{TAMUdust2020} database to reproduce the scattering behavior of natural dust samples has been validated through direct comparison with experimental measurements \citep{Martikainen2025MNRAS.537.1489M}. 
During this work, we adopt a sphericity $\Psi=0.695$, corresponding to the most irregular morphology in the database. We refer the reader to Figure 1 of \cite{Saito2021JAtS...78.2089S} for a clear reference on how the grain morphology changes as sphericity is modified.
We use the same size distribution, density, and refractive index as for the spherical grains to ensure consistent comparison.

The \texttt{TAMUdust2020} database provides quantities such as volume, projected-surface area, extinction efficiency, and albedo. For irregular grains, the equivalent radius $a$ can be defined in two ways: (i) as the radius of a sphere with the same projected-surface area averaged over all orientations, or (ii) as the radius of a sphere with the same volume. The opacities obtained for both definitions of the radius are listed in the Table \ref{tab:opacities}.
Regardless of the adopted definition, the solid irregular grains exhibit stronger scattering by factors of $\sim 1.5$ (area-based) to $\sim 2.5$ (volume-based) than solid spheres, consistent with previous numerical and laboratory studies that show enhanced scattering in irregular grains \citep[e.g.,][]{Kirchschlager2020, Lin2023}. Throughout this article, we will use the definition of radius based on the projected-surface area. 

Once the definition of $a$ is fixed, the mass extinction opacity is calculated as
\begin{equation}
    k_\mathrm{ext} = \frac{\int Q_\mathrm{ext}(a,\lambda)\,\pi a^2\,n(a)\,da}{\int \frac{4}{3}\pi a^3 \rho_\mathrm{dust}\,n(a)\,da},
\end{equation}
where $Q_\mathrm{ext}$ denotes the extinction efficiency, $n(a)$ represents the size distribution, and $\rho_\mathrm{dust}$ indicates the dust density. Similar expressions are employed for $k_\mathrm{sca}$ and $k_\mathrm{abs}$.

\begin{figure*}[ht]
    \centering
    \includegraphics[width=\linewidth]{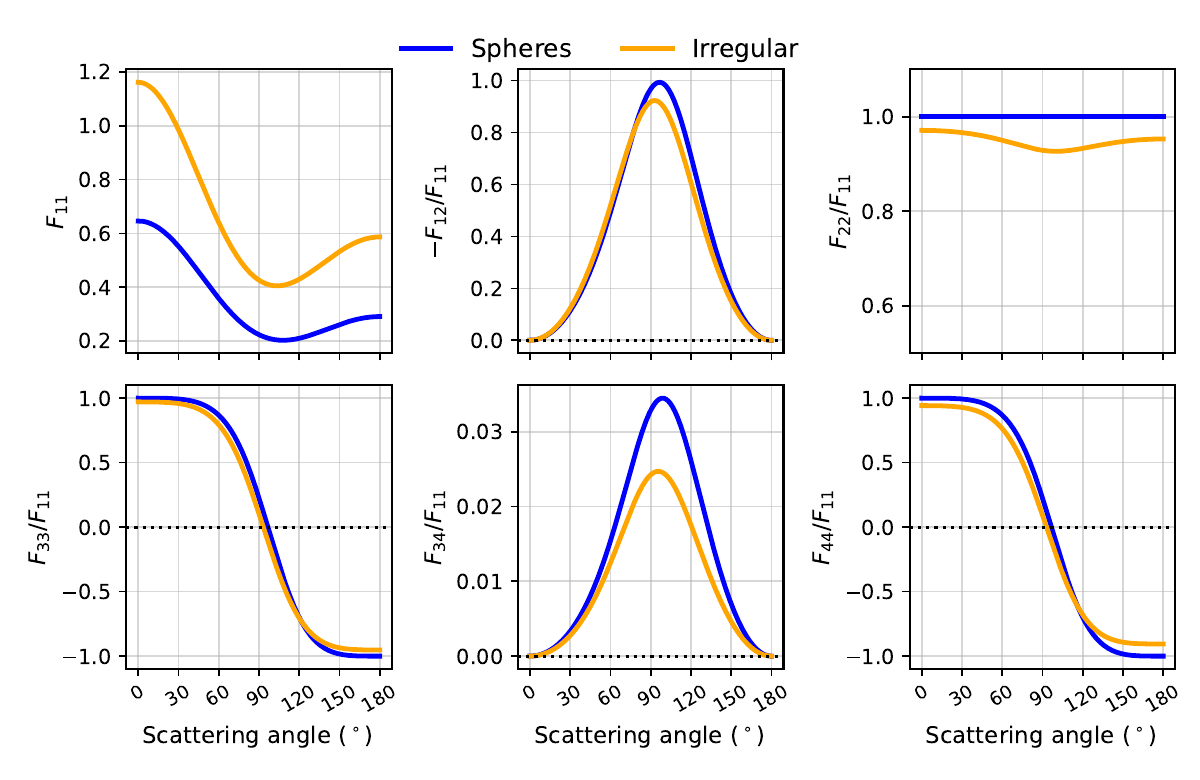}
    \caption{Computed scattering matrix elements for a size distribution, with a $a_\mathrm{max} = 160~\mu$m, composed of spherical particles (blue) and irregular particles (orange) at a wavelength of 1 mm. In both cases, the complex refractive index is equal to $m = 2.2995+i\,0.02031$.} 
    \label{fig:scattering_matrices_amax160}
\end{figure*}

The comparison between spherical and irregular grains can be illustrated through the elements of their scattering matrices \citep{Mishchenko2000}. For unpolarized incident light asuming randomly oriented, 
mirror-symmetric particles \citep[e.g.,][]{Hovenier2004ASSL}, the $F_{11}$ element provides the total scattered intensity, while $-F_{12}/F_{11}$ defines the degree of linear polarization. The ratio $F_{22}/F_{11}$ tracks the modification of linear polarization, either parallel or perpendicular to the scattering plane, with deviations from unity indicating non-sphericity. Similarly, $F_{33}/F_{11}$ describes polarization at $\pm45^\circ$, while $F_{34}/F_{11}$ and $F_{44}/F_{11}$ encode the conversion between linear and circular polarization and the conservation of circular polarization, respectively. Figure \ref{fig:scattering_matrices_amax160} shows the scattering matrix for the two dust samples. We can see that irregular grains enhance both forward and backward scattering relative to spheres, producing systematic deviations in $F_{22}/F_{11}$, even when most particles are smaller than the wavelength.

\subsection{Radiative Transfer Models}
To investigate the emergent polarized emission, we constructed a set of parametric disk models that span three optical depth regimes: a fully optically thick disk, a fully optically thin disk, and a hybrid model with an optically thick center and optically thin outskirts. The objective of this optical depth exploration is to determine how the polarization pattern changes with optical depth. In optically thick regions, photons are predominantly scattered by nearby grains, whereas in optically thin regions, they can originate from distant parts of the disk before being scattered.

In all our models, the dust surface density follows the similarity solution of \citet{Lynden-Bell1974MNRAS.168..603L},
\begin{equation}
    \Sigma(R) = \Sigma_0 \left(\frac{R}{R_0}\right)^{-\gamma} \exp\left[-\left(\frac{R}{R_0}\right)^{2-\gamma}\right],
\end{equation}
where $\Sigma_0 = \tau_0/\kappa_\mathrm{ext}$ and $\tau$ represent the optical depth. The parameters adopted for each optical regime are listed in Table \ref{tab:model_param_table}. To construct the disks, we use the $\kappa_\mathrm{ext}$ of the spheres. This approach enables us to establish the masses of the disk models.

\begin{table}
    \centering
    \begin{tabular}{lcccc}
        \hline
        Model & $R_0$ (au) & $\tau_{0}$ & $\gamma$ & Mass ($\mathrm{M_\odot}$) \\
        \hline
        Optically thick & 50 & 100   & 0.2  & $1.39\times10^{-3}$ \\
        Optically thin  & 50 & 10  & 0.2 &  $1.39\times10^{-4}$ \\
        Hybrid          & 40 & 6.0   & 0.2 &  $3.61\times10^{-5}$ \\
        \hline
    \end{tabular}
    \caption{Parameters adopted for the disk models in different optical depth regimes.}
    \label{tab:model_param_table}
\end{table}

The vertical dust density profile is assumed to follow a Gaussian distribution,
\begin{equation}
    \rho(R,z) = \frac{\Sigma(R)}{H(R)\sqrt{2\pi}} \exp\left[-\frac{1}{2}\left(\frac{z}{H(R)}\right)^2\right],
\end{equation}
with a scale height described by
\begin{equation}
    H(R) = H_0 \left(\frac{R}{R_0}\right)^{1.25}, \qquad H_0/R_0 = 0.01.
\end{equation}

The disks are vertically isothermal and follow a radial temperature profile $T(R)=T_0 (R/R_0)^{-0.5}$, normalized to $T_0=30$ K at 50 au, consistent with passive disk models \citep{Chiang1997ApJ...490..368C, Stephens2023}. The outer radius is set to 100 au. For the purposes of this work, we do not take into account the settling of dust; rather, we assume that the optical properties of the dust are uniform across the entire disk. This parametric setup provides a controlled framework for isolating the effects of dust shape and optical depth on the resulting polarization patterns.

Radiative transfer simulations were performed using the \texttt{RADMC-3D} code. 
For each optical depth regime, the disk geometry, temperature structure, and grain size distribution were kept identical.
The simulations were carried out at a wavelength of 1 mm, where the dust grains with a maximum size of $a_{max}=160~\mathrm{\mu m}$ maximize the self-scattering.
The full Stokes parameters $(I, Q, U, V)$ were computed using $10^9$ photon packets to ensure a statistically robust sampling of the scattered radiation field. The synthetic images have $1000 \times 1000$ pixels and a total field of view of 250 au, which is large enough to encompass the entire disk emission.

To quantify polarization, we define the polarized fraction as $p_{f} \equiv PI/I$, where $PI = \sqrt{Q^2+U^2}$ is the polarized intensity. 

\section{Results and Discussion}\label{sec:results_discussion}

\begin{figure*}
    \centering
    \includegraphics[width=0.9\linewidth, height=0.9\textheight]{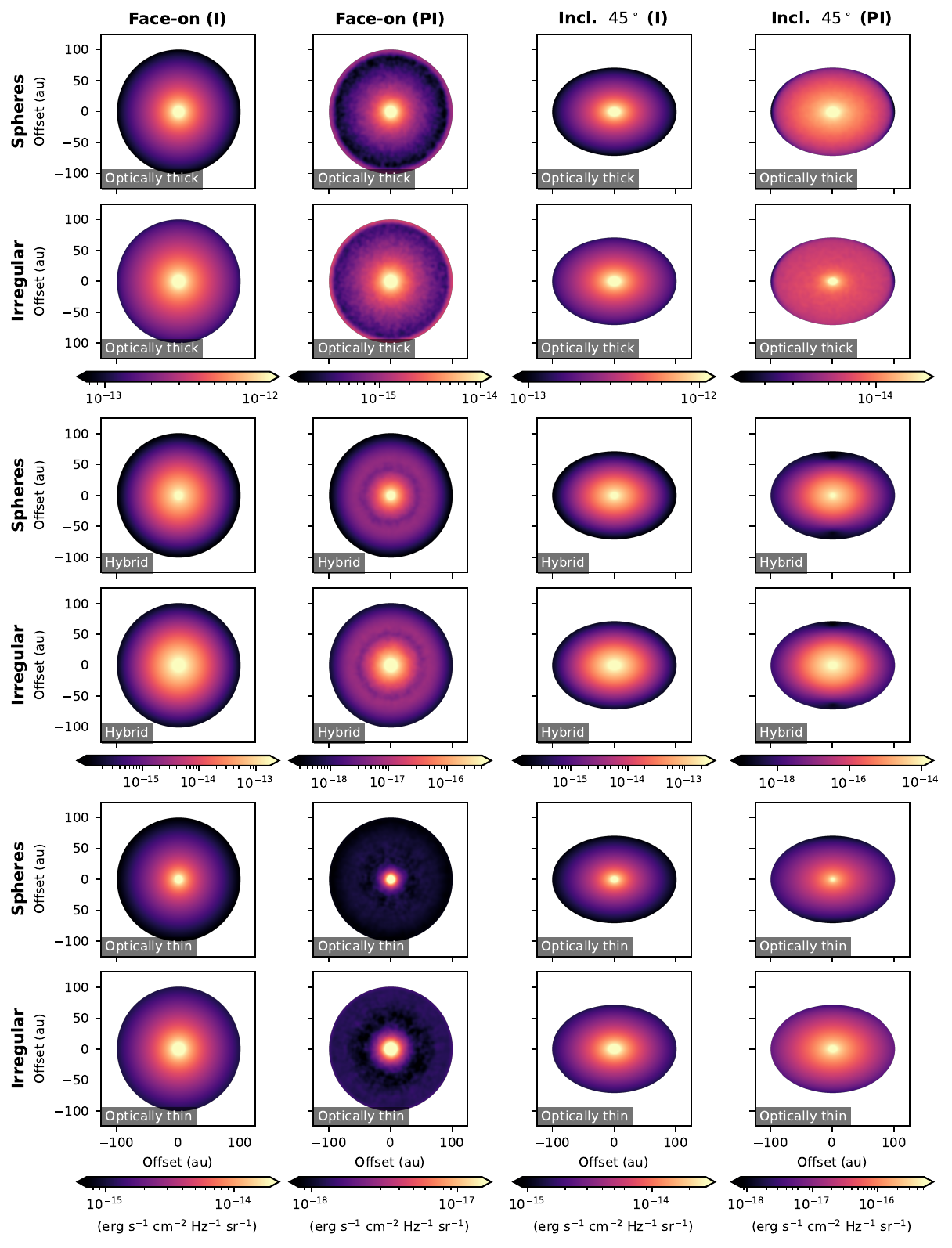}
    \caption{Total and polarized intensity maps for the disk models with different grain geometries at 1 mm. The first and third columns show the total intensity ($I$) for the models with spherical (top) and hexahedral (bottom) grains in the face-on and inclined ($45^\circ$) views, respectively. The second and fourth columns display the corresponding polarized intensity ($PI$). Both grain size distribution has a $a_\mathrm{max} = 160~\mu$m, utilize the DSHARP compositions, and share the complex refractive index m = 2.2995 + i 0.02031. }
    \label{fig:IandP_plots}
\end{figure*}

Figure \ref{fig:IandP_plots} illustrates the resulting total intensity ($I$; columns 1 and 3) and polarized intensity ($PI$, columns 2 and 4) maps at 1 mm for disk models featuring spherical and irregular dust grains, shown for both face-on and 45 $^\circ$ inclined view angles. Comparing the emission maps across the different optical depth regimes (optically thick, hybrid, and optically thin), we observe that irregular grains (bottom rows of each pair) consistently produce a more extended emission profile than their spherical counterparts (top rows). This result is particularly evident in the $PI$ maps where irregular particles maintain detectable signals at larger radial offsets. This extended emission comes from the combined effect of higher absorption and scattering opacities from irregular grains.

\begin{figure*}[ht]
    \centering
    \includegraphics[width=\linewidth,trim={0cm 0cm 0cm 0.6cm},clip]{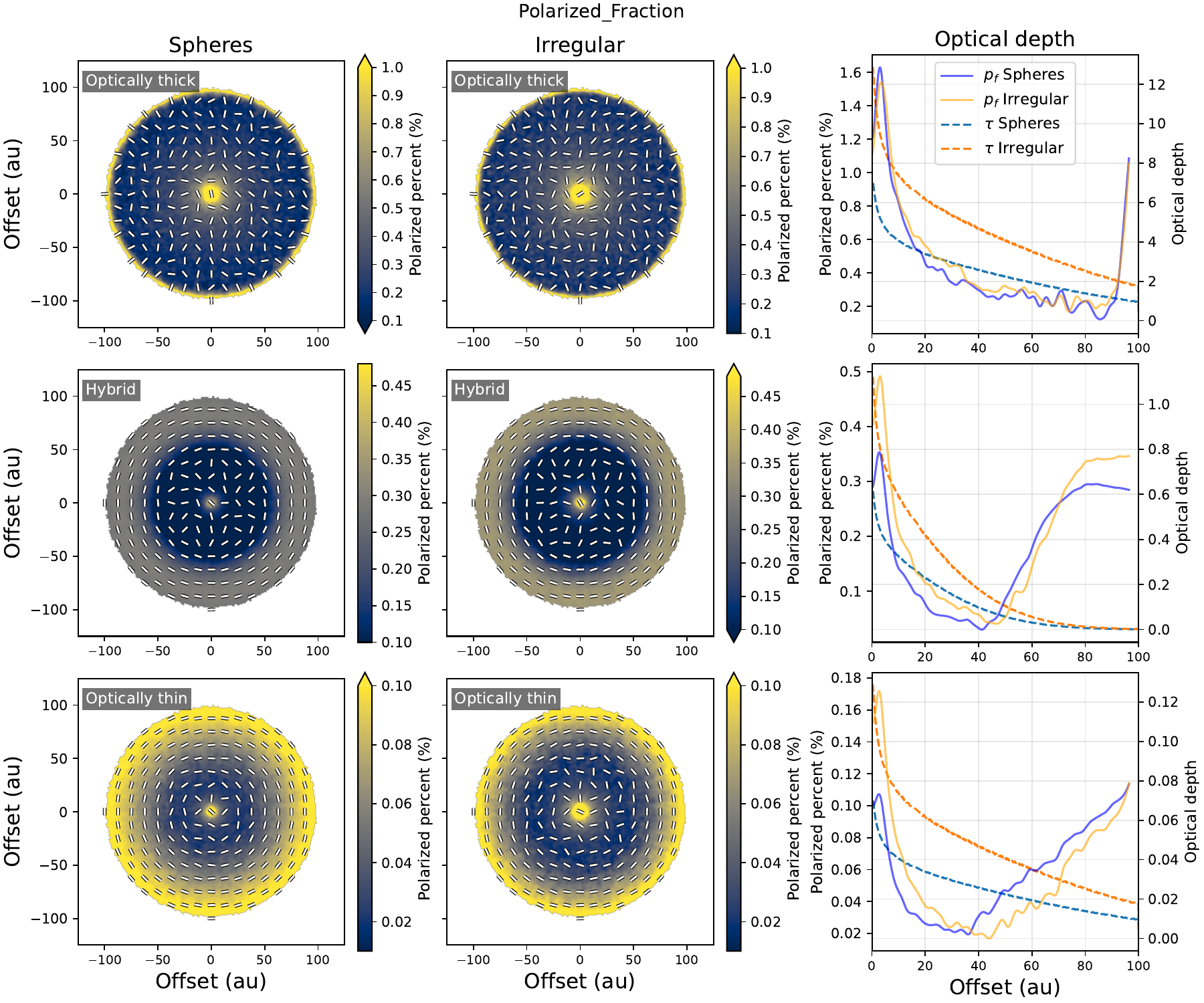} 
    \caption{Polarized fraction maps at 1 mm are shown for the optically thick (upper row), hybrid (middle row), and optically thin (bottom row) disk models.
    The left column displays maps for models using a size distribution of spherical grains, while the center column shows models using irregular grains. Both grain size distribution has a $a_\mathrm{max} = 160~\mu$m, utilize the DSHARP compositions and share the complex refractive index $m = 2.2995 + i\,0.02031$.
    The right column presents the polarized fraction (solid lines) and optical depth (dashed lines) profiles across the $x$-axis for each disk model, with spheres shown in blue and irregular grains in orange.}
    \label{fig:all_models_wav1mm_faceon}
\end{figure*}

Figure \ref{fig:all_models_wav1mm_faceon} shows the polarization fraction maps for the optically thick (top row), hybrid (middle row), and optically thin (bottom row) disk models viewed face-on.
The left and central columns correspond to models with spherical and irregular grains, respectively, while the right panels illustrate the radial profiles of the polarized fractions $p_f$ and the optical depth $\tau$.

In the optically thick model, polarization is concentrated in the outer radial regions, where the radiation field is more anisotropic. In these regions, the polarization vectors are radially aligned, as most of the scattered radiation originates from the dust distributed along the edge of the disk.
Toward the inner regions, the polarization diminishes because the local radiation field is nearly isotropic, resulting in the cancelation of the polarization. The few residual polarized components that remain exhibit polarized vectors that are randomly oriented. 

In the hybrid model, where the optical depth decreases radially more rapidly from an opaque center to an optically thin outskirt, the polarization vectors exhibit an azimuthal aligned pattern across the disk.
In the optically thick zone, irregular grains produce slightly higher polarization fractions than spherical grains. At the transition between optical regimes ($R\sim40$ au), the polarization fraction reaches a minimum and subsequently increases outward. In the outer optically thin region, the polarization vectors are azimuthally aligned due to the anisotropy of the radiation field arising from photons from the inner regions.

In the optically thin case, both grain models display low polarization fractions ($p_f \sim 0.1\%$) with a uniformly azimuthal vector orientation.The polarization fraction increases with the radius, reaching its highest value at the edges of the disk, where the anisotropy of the radiation field is greater.

Overall, in face-on geometry, irregular grains consistently produce a slightly higher peak maximum polarization fraction across all optical depth regimes.
An additional effect of using irregular solid grains instead of spherical grains is an increase in optical depth, which results from their higher extinction, primarily due to enhanced scattering.  

\begin{figure*}[ht]
    \centering
    \includegraphics[width=\linewidth,trim={0cm 0cm 0cm 0.6cm},clip]{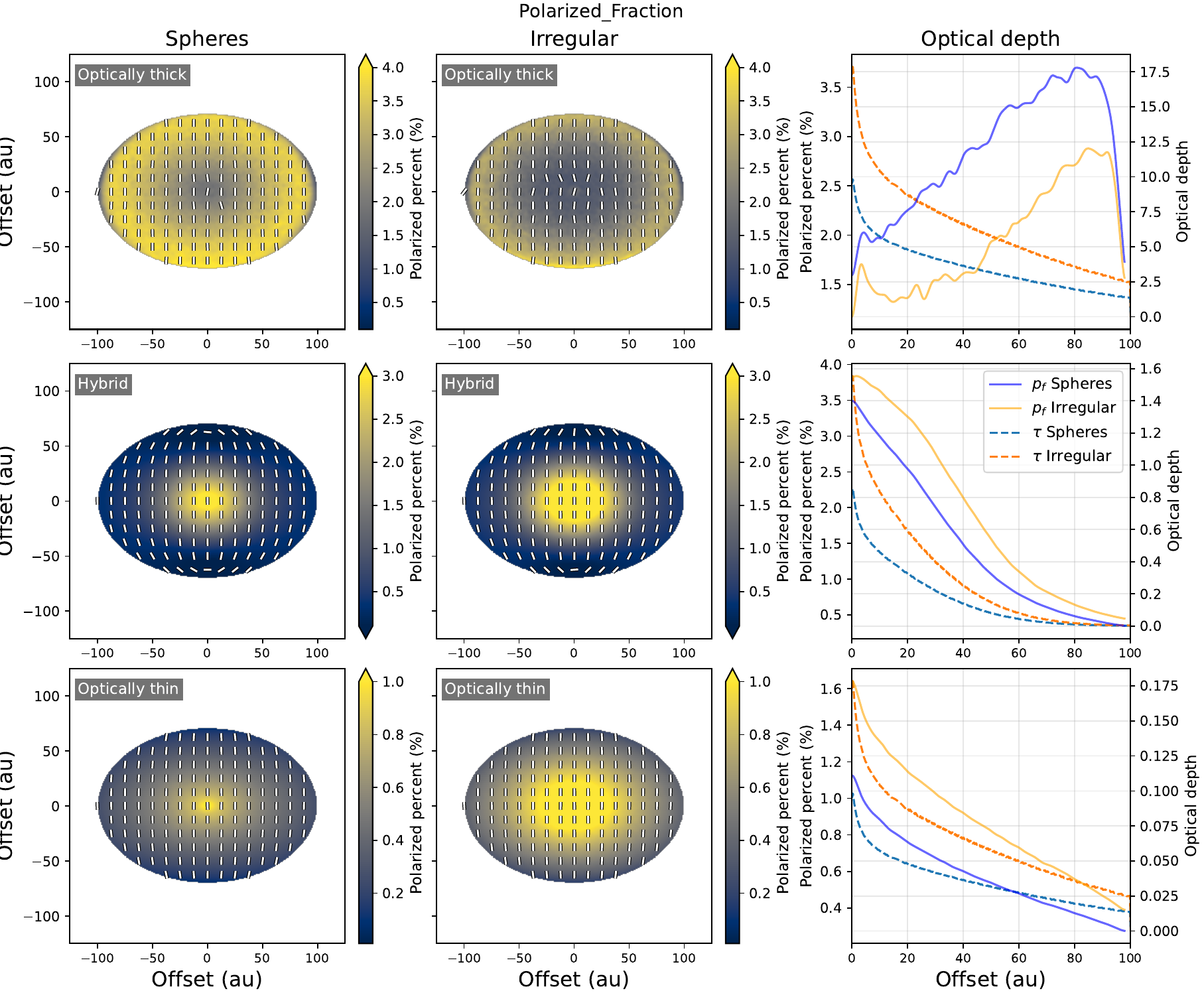} 
    \caption{Same as a Figure \ref{fig:all_models_wav1mm_faceon} for a view with inclination of $45^{\circ}$. }
    \label{fig:all_models_wav1mm_incl45}
\end{figure*}

Figure \ref{fig:all_models_wav1mm_incl45} presents the same set of models viewed at an inclination of $45^\circ$. In the optically thick disk, the polarization vectors align predominantly with the disk minor axis, consistent with the expected scattering geometry in inclined, optically thick disks \citep{Kataoka2016a}. The polarization fractions increase from $\sim 1\%$ near the center to $4\%$ toward $R\sim80$ au for spherical grains, while the irregular grains show slightly lower values ($\sim 3 \%$).

In the hybrid model, both grain populations exhibit a smooth radial decline in the polarized fraction reaching $\sim 3\%$ at intermediate radii and $<1\%$ beyond $R\sim80$ au. The overall morphology remains consistent across grain models, but, unlike the optically thick disk, spherical grains yield marginally weaker polarization across the disk.

In the optically thin model, the vectors maintain a uniform orientation parallel to the disk's minor axis, with maximum $p_f \sim 1.5\%$. The polarization decreases steadily with radius, tracking the optical depth profile. 

Comparing both inclinations, the polarization fraction clearly increases with viewing angle, especially for the optically thick disk, where the projected anisotropy of the radiation field enhances the net scattering-induced polarization. The transition from azimuthal to minor-axis alignment between the face-on and inclined views also follows the expected behavior for self-scattering, as reported in previous theoretical and observational studies of HL Tau and IM Lup \citep[e.g.,][]{Kataoka2016b, Hull2018ApJ...860...82H, Harrison2019ApJ...877L...2H, Lin2024}.

Across all regimes, irregular grains yield polarization fractions slightly higher than those of spherical grains, except for the optically thick model viewed at $45^\circ$, where the higher scattering of the irregular grains reduces the total polarization. 

In our models, we observe that when we represent the disk dust as irregular solid particles, the optical depth is consistently higher than when we model it as spherical solid particles.
This implies that, when we measure the brightness of a disk, assuming it is composed of irregular particles will require less mass than modeling it as spheres.

\subsection{ Polarization of large millimetre-sized grains}

Previous studies \citep[e.g.,][]{Lin2023} have demonstrated that the maximum grain size strongly influences the polarization characteristics of disk emission at millimeter wavelengths. As grain size approaches or exceeds the observing wavelength, their scattering phase functions and polarization efficiencies change significantly, often leading to a reversal in the polarization orientation \citep{Kataoka2015, Yang2016a}. For spherical grains, this transition typically occurs near the characteristic size parameter $a_\mathrm{max} \sim \lambda/2\pi$. In contrast, non-spherical and aggregate grains tend to exhibit smoother transitions and can maintain significant polarization levels even when their sizes become comparable to or larger than the observing wavelength \citep{Kirchschlager2019, Kirchschlager2020}.

Motivated by these results, we explore how increasing $a_\mathrm{max}$ modifies the scattering properties and resulting polarization patterns in our models when using irregular grains instead of spheres. We analyzed the dust size distributions from maximum grain sizes from $a_\mathrm{max} = 10$ to $a_\mathrm{max}=10^4$, focusing on representative cases that correspond to substantial changes in the scattering regime. Next, we describe the properties of the size distribution with $a_\mathrm{max} = 300$, $500$, and $1000~\mu$m, corresponding to size parameters of $\sim 2, 3, 6$, which are representative of large changes in the properties of the dust samples. 

The first effect observed is that as the maximum grain size increases, both absorption and scattering opacity initially increase, leading to a substantial increase in total extinction relative to smaller-grain distributions (see Table \ref{tab:opacities} for quantitative values). This behavior reflects the increasing contribution of larger particles to the overall cross section. Although both spherical and irregular grains follow this general trend, irregular particles systematically exhibit higher extinction opacities and a stronger dependence on grain size.

When the maximum grain size approaches values comparable to the observing wavelength, the behavior begins to diverge between the two morphologies. For spherical grains, the opacities tend to saturate and vary only slightly with further increases in size. In contrast, irregular grains show a more pronounced evolution, particularly in their scattering properties, resulting in a steeper variation of total extinction with grain size.

For even larger size distributions, the scattering opacity gradually decreases while absorption continues to grow, although more slowly. This reflects the transition from the Rayleigh to the Mie regime, where forward scattering becomes increasingly dominant and the polarization efficiency is reduced. In all sizes explored, irregular grains maintain higher effective absorption and display a stronger sensitivity to size evolution than spherical grains, consistent with laboratory and numerical studies of irregular particles \citep[e.g.,][]{Kirchschlager2020, Lin2023}.

An important characteristic of scattering-induced polarization is that the polarization angle can reverse as the maximum grain size increases and the scattering regime transitions from Rayleigh to Mie. This phenomenon, known as polarization reversal, occurs when the direction of maximum polarization changes relative to the local radiation anisotropy \citep{Daniel1980, Kataoka2015, Yang2016a}. The inversion arises from the angular dependence of the scattered intensity, which is governed by the elements of the scattering matrix $F_{11}$ and $F_{12}$. In particular, the sign of the ratio $F_{12}/F_{11}$ determines the orientation of the scattered polarization.

\begin{figure}[ht]
    \centering
    \includegraphics[width=\linewidth]{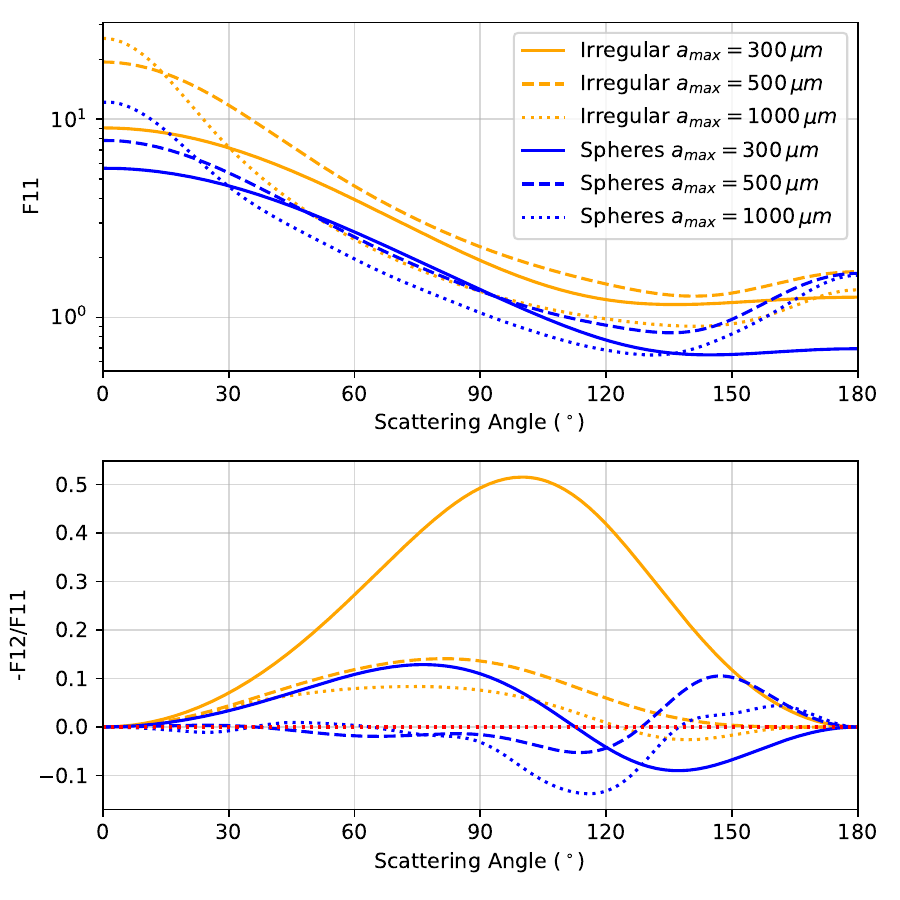}
    \caption{Scattering matrix elements for spherical (blue) and irregular (orange) grains with $a_\mathrm{max}=300$ (solid), $500$ (dashed), and $1000~\mu\mathrm{m}$ (dotted).}
     \label{fig:zoom_scattering_matrix_large_grains}
\end{figure}

Figure \ref{fig:zoom_scattering_matrix_large_grains} shows the scattering matrix elements for the dust samples with $a_\mathrm{max} = 300$, $500$, and $1000~\mu\mathrm{m}$. In the upper panel, we observe that as $a_\mathrm{max}$ increases, the scattering becomes increasingly anisotropic. Both spherical and irregular grains show enhanced forward scattering, but the effect is much stronger for the irregular grains, which also exhibit higher overall scattering efficiency.

The lower panel of Figure \ref{fig:zoom_scattering_matrix_large_grains} presents the corresponding polarization ratio $-F_{12}/F_{11}$. For the distribution with $a_\mathrm{max} = 300~\mu\mathrm{m}$, spherical grains display a polarization peak of about 15\% at scattering angles near $70^{\circ}$, followed by a change in sign between $110^{\circ}$ and $175^{\circ}$, indicating the onset of polarization reversal. In contrast, the irregular grains maintain a positive polarization ratio, with a broad maximum near $90^{\circ}$, but with a reduced efficiency of roughly 50\% relative to the models with dust size distribution with $a_\mathrm{max}\sim\lambda/2\pi$. When $a_\mathrm{max}$ is increased to $500~\mu\mathrm{m}$, the irregular particles continue to show predominantly positive polarization, though with lower amplitude, while spherical grains exhibit an extended region of negative polarization between $0^{\circ}$ and $125^{\circ}$, where the net polarization remains near zero. At \(a_\mathrm{max}=1000~\mu\mathrm{m}\), the spherical-grain polarization efficiency increases slightly but remains negative over most scattering angles, whereas the irregular grains still retain positive polarization for the majority of directions.

These results demonstrate that the morphology of dust grains plays a crucial role in the angular dependence of scattered polarization. Irregular particles, which lack the symmetric interference patterns characteristic of Mie spheres, exhibit smoother transitions and retain positive polarization over a wider range of sizes. Consequently, models that incorporate irregular grains can reproduce the observed polarization levels in protoplanetary disks while allowing for maximum grain sizes up to at least twice as large as those inferred under the spherical-grain approximation.

\begin{figure}[ht]
    \centering
    \includegraphics[width=\linewidth]{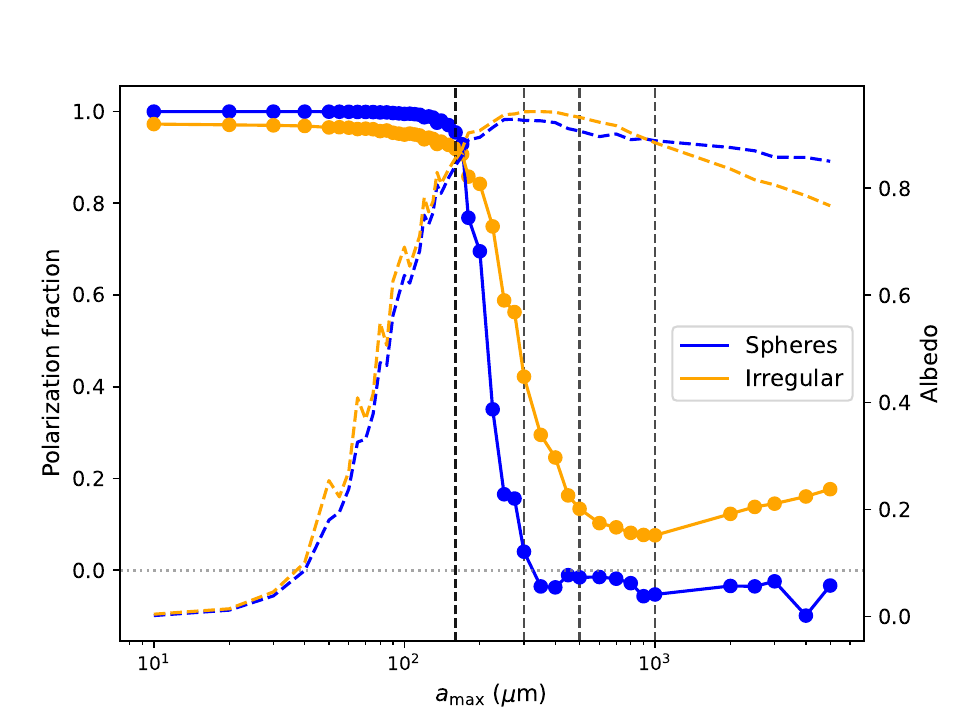}
    \caption{ Polarization fraction (solid) and albedo (dashed) as a function of $a_\mathrm{max}$ for spherical (blue) and irregular (orange) grains at an angle of $90^\circ$.
    Spherical grains lose polarization efficiency beyond $a_\mathrm{max} \approx 300~\mu\mathrm{m}$, while irregular grains maintain detectable polarization up to millimeter sizes despite similar albedo trends. Dashed vertical lines indicate the maximum grain sizes used in the analysis 
($a_{\max} = 160,300, 500,$ and $1000\,\mu\mathrm{m}$).}
    \label{fig:pol_fraccVSamax}
\end{figure}

Figure \ref{fig:pol_fraccVSamax} further quantifies these trends by showing the dependence of the polarization fraction and albedo on $a_\mathrm{max}$, at an angle of $90^\circ$. The polarization fraction (solid lines) decreases sharply once the grains grow beyond $\lambda/2\pi$, while the albedo (dashed lines) continues to increase up to $a_\mathrm{max}\sim300~\mu\mathrm{m}$ before flattening. For spherical grains, the polarization efficiency drops nearly to zero beyond $a_\mathrm{max}\sim300~\mu\mathrm{m}$, reflecting the dominance of forward scattering and polarization reversal. In contrast, irregular grains maintain polarization fractions above 20\% even for $a_\mathrm{max}\sim1000~\mu\mathrm{m}$, despite a modest decline in albedo. This confirms that irregular particles can produce detectable polarization signatures at millimeter wavelengths, consistent with the values observed in bright disks such as HL~Tau and IM~Lup \citep{Kataoka2017, Hull2018ApJ...860...82H, Harrison2024}.

To examine the impact of increasing maximum grain size, we performed radiative transfer simulations at 1~mm using dust size distributions spanning $a_\mathrm{max}=300$, 500, and 1000~$\mu$m. These values were selected to sample the transition from size parameter $\sim 1$ to $>1$ at millimeter wavelengths. For each grain size, we computed models including multiple scattering under three representative optical depth regimes: optically thick, hybrid, and optically thin. The most representative case, in which we still observe high polarization fraction levels and a significant difference between spherical and irregular particles, begins to emerge when we use a size distribution with $a_\mathrm{max} = 500~\mu\mathrm{m}$. At this $a_\mathbf{max}$, the polarization reversal becomes evident, as it is where the transition from the Rayleigh to the Mie regime occurs ($x = 2\pi a / \lambda \sim 1$), which causes the change in the sign of the relation $-F_{12}/F_{11}$ and, consequently, in the direction of the polarization vectors \citep{Kataoka2015, Yang2016a}.

\begin{figure*}[ht]
    \centering
    \includegraphics[width=\linewidth,trim={0cm 0cm 0cm 1cm},clip]{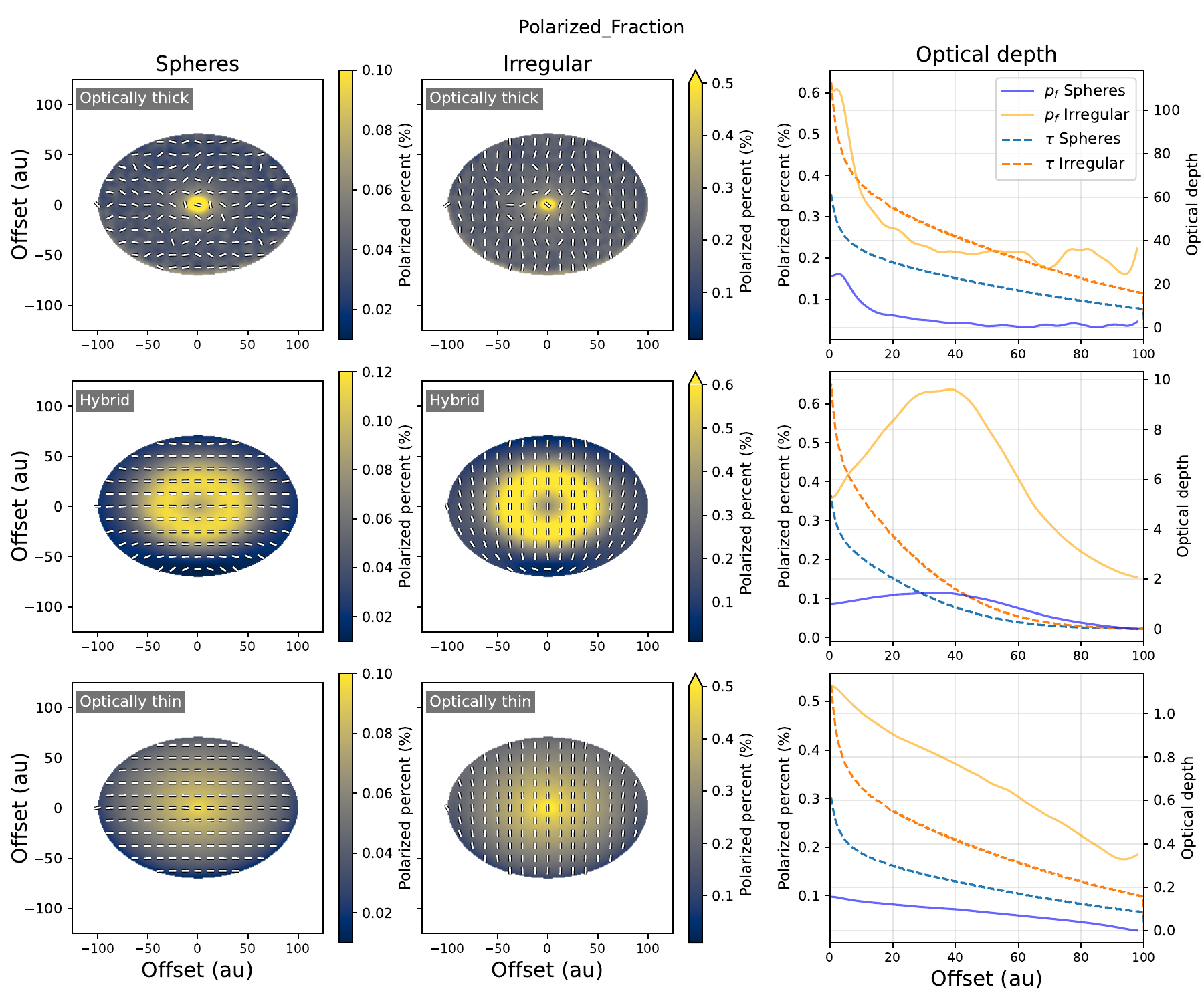}
    \caption{Polarized fraction maps at 1~mm for models with $a_\mathrm{max} = 500\,\mu$m and disk inclination $i=45^\circ$.
    Rows correspond to the optically thick, hybrid, and optically thin disk models, respectively.
    The left column displays maps for models using spherical grains, while the center column shows models using irregular grains. Both grain size distribution has a $a_\mathrm{max} =500~\mu$m, utilize the DSHARP compositions and share the complex refractive index $m = 2.2995 + i\,0.02031$.
    The right column presents the polarized fraction (solid lines) and optical depth (dashed lines) profiles across the $x$-axis for each disk model, with spheres shown in blue and irregular grains in orange.}
    \label{fig:polfraction_500um_incl45}
\end{figure*}

Figure \ref{fig:polfraction_500um_incl45} shows the polarization fraction maps at 1~mm for $a_\mathrm{max}=500\,\mu$m and an inclination of $45^\circ$. The left column corresponds to spherical grains, the middle column to irregular grains, and the rows represent the optically thick, hybrid, and optically thin disk models. 

Across all optical-depth regimes, irregular grains systematically produce higher polarization fractions than spherical grains. The spherical models remain near or below the $\sim0.1$--$0.12$\% level, whereas the irregular models reach values up to $\sim0.5$\%, reflecting their enhanced scattering efficiency at this wavelength.

In the hybrid model, the polarization fraction displays a pronounced radial maximum at $R \sim 50$~au, forming a ring-like structure. This feature arises from the interplay between the centrally concentrated total intensity and the more spatially extended polarized intensity. In the inner disk, the high optical depth suppresses the polarization fraction despite strong total emission. At intermediate radii, where the optical depth becomes moderate and the anisotropy of the radiation field is strongest, scattering becomes most efficient and the polarization fraction reaches a maximum. The ring therefore traces the transition between optically thick and optically thin conditions.

Regarding the polarization orientation, a clear morphological difference emerges between spherical and irregular grains in the hybrid and optically thin regimes. For spherical grains, the polarization vectors are predominantly aligned parallel to the projected major axis (horizontal direction), with deviations only near the outer disk edges. In contrast, for irregular grains, the vectors remain largely parallel to the projected minor axis (vertical direction) across most of the disk, again with minor deviations near the outer boundaries. 

In the optically thick regime, the polarization signal is weaker and the radiation field more isotropic, making the orientation differences less pronounced. However, as the optical depth decreases, the systematic reorientation between the two morphologies becomes evident across the majority of the disk, reflecting their different size-dependent scattering behavior at $a_\mathrm{max}=500\,\mu$m.

\section{Conclusions}
\label{sec:conclusions}

We have investigated the polarized emission produced by dust self-scattering in protoplanetary disks using radiative transfer simulations that incorporate both solid spherical and solid irregular dust grains. The optical properties of the irregular grains were obtained from the \texttt{TAMUdust2020} database, which is based on hexahedral-shaped particles that reproduce the laboratory-measured scattering behavior of realistic particles. Our models span three optical depth regimes—optically thick, optically thin, and an intermediate case—to assess how grain morphology influences polarization patterns and their dependence on disk geometry.

We find that spherical and  irregular grains produce broadly similar polarization fractions and opacity behaviors across the explored regimes. In face-on, optically thin configurations, polarization vectors remain predominantly azimuthal due to the anisotropy of the radiation field, while at higher optical depths this preferential alignment weakens as the radiation field becomes more isotropic. In inclined disks, both grain morphologies produce polarization vectors aligned with the projected minor axis, consistent with observed scattering-induced polarization patterns.

The most notable difference between spherical and irregular grains concerns the critical grain size at which the polarization orientation reverses. For spherical grains, this flip occurs as soon as twice the size parameter at which the polarization is maximum, as predicted by Mie theory. In contrast, irregular grains shift this transition to larger effective size parameters and with smallest intensity, suppressing the polarization reversal for grains moderately larger than the observing wavelength. This behavior is consistent with laboratory and numerical studies of irregular particles and highlights a potential bias that may arise when interpreting polarization morphology under the spherical assumption.

Irregular grains also exhibit moderately higher extinction efficiencies at millimeter wavelengths, primarily due to increased scattering opacities. However, the resulting differences in inferred grain sizes and surface dust densities are still defined by the smallest particles in the distribution. Therefore, adopting solid irregular grains instead of solid spheres does not significantly alter grain size estimates from intensity maps.

These results indicate that grain irregularity alone does not resolve the long-standing discrepancy between grain sizes inferred from spectral index analyses and those inferred from polarization studies. Rather, our work demonstrates that incorporating realistic grain morphologies is a necessary step toward more physically motivated disk models. More substantial effects may arise from additional grain properties, such as porosity or aggregate structure, which modify both the effective refractive index and scattering phase function. A detailed exploration of porous and aggregate grains is left for future work.

Overall, this study advances the modeling of polarized emission in protoplanetary disks by moving beyond the idealized spherical assumption and quantifying the impact of grain morphology in a controlled manner. Continued efforts incorporating multi-wavelength simulations and more complex grain structures will be essential for refining constraints on dust properties during the early stages of planet formation.

\begin{acknowledgments}
We appreciate all comments from the anonymous referee
that helped improve our paper.
J.M.J.-D. gratefully acknowledges the financial support provided by the SECIHTI through the Estancias Posdoctorales por Mexico CVU 864232.
C.C.-G. and J.M.J.-D. acknowledges support from UNAM DGAPA-PAPIIT grant IG101224.
J.M.J.-D., C.C.-G., D.G., O.M., G.V. were funded by the IAA-CSIC Severo Ochoa excellence award CEX2021-001131-S through its incoming visitor program and grant PID2024-156713OB-I00/AEI/FEDER funded by MCIN/AEI/10.13039/501100011033. This work benefited from the UNAM-NRAO Memorandum of Understanding in the framework of the Next Generation Very Large Array (ngVLA) Project (MOU-UNAM-NRAO-2023).
\end{acknowledgments}

\software{astropy \citep{astropy:2013, astropy:2018, astropy:2022}, TAMUdust2020 \citep{Saito2021JAtS...78.2089S},Optool \citep{Dominik2021ascl.soft04010D}, RADMC-3D \citep{Dullemond2012ascl.soft02015D}. }

\bibliography{bibliography}{}

@ARTICLE{ALMA_Brogan2015,
       author = {{ALMA Partnership} and {Brogan}, C.~L. and {P{\'e}rez}, L.~M. and {Hunter}, T.~R. and {Dent}, W.~R.~F. and {Hales}, A.~S. and {Hills}, R.~E. and {Corder}, S. and {Fomalont}, E.~B. and {Vlahakis}, C. and {Asaki}, Y. and {Barkats}, D. and {Hirota}, A. and {Hodge}, J.~A. and {Impellizzeri}, C.~M.~V. and {Kneissl}, R. and {Liuzzo}, E. and {Lucas}, R. and {Marcelino}, N. and {Matsushita}, S. and {Nakanishi}, K. and {Phillips}, N. and {Richards}, A.~M.~S. and {Toledo}, I. and {Aladro}, R. and {Broguiere}, D. and {Cortes}, J.~R. and {Cortes}, P.~C. and {Espada}, D. and {Galarza}, F. and {Garcia-Appadoo}, D. and {Guzman-Ramirez}, L. and {Humphreys}, E.~M. and {Jung}, T. and {Kameno}, S. and {Laing}, R.~A. and {Leon}, S. and {Marconi}, G. and {Mignano}, A. and {Nikolic}, B. and {Nyman}, L. -A. and {Radiszcz}, M. and {Remijan}, A. and {Rod{\'o}n}, J.~A. and {Sawada}, T. and {Takahashi}, S. and {Tilanus}, R.~P.~J. and {Vila Vilaro}, B. and {Watson}, L.~C. and {Wiklind}, T. and {Akiyama}, E. and {Chapillon}, E. and {de Gregorio-Monsalvo}, I. and {Di Francesco}, J. and {Gueth}, F. and {Kawamura}, A. and {Lee}, C. -F. and {Nguyen Luong}, Q. and {Mangum}, J. and {Pietu}, V. and {Sanhueza}, P. and {Saigo}, K. and {Takakuwa}, S. and {Ubach}, C. and {van Kempen}, T. and {Wootten}, A. and {Castro-Carrizo}, A. and {Francke}, H. and {Gallardo}, J. and {Garcia}, J. and {Gonzalez}, S. and {Hill}, T. and {Kaminski}, T. and {Kurono}, Y. and {Liu}, H. -Y. and {Lopez}, C. and {Morales}, F. and {Plarre}, K. and {Schieven}, G. and {Testi}, L. and {Videla}, L. and {Villard}, E. and {Andreani}, P. and {Hibbard}, J.~E. and {Tatematsu}, K.},
        title = "{The 2014 ALMA Long Baseline Campaign: First Results from High Angular Resolution Observations toward the HL Tau Region}",
      journal = {\apjl},
         year = 2015,
        month = jul,
       volume = {808},
       number = {1},
          eid = {L3},
        pages = {L3},
          doi = {10.1088/2041-8205/808/1/L3},
archivePrefix = {arXiv},
       eprint = {1503.02649},
 primaryClass = {astro-ph.SR},
       adsurl = {https://ui.adsabs.harvard.edu/abs/2015ApJ...808L...3A}
}

@ARTICLE{Andrews2018DSHARP_I,
       author = {{Andrews}, Sean M. and {Huang}, Jane and {P{\'e}rez}, Laura M. and {Isella}, Andrea and {Dullemond}, Cornelis P. and {Kurtovic}, Nicol{\'a}s T. and {Guzm{\'a}n}, Viviana V. and {Carpenter}, John M. and {Wilner}, David J. and {Zhang}, Shangjia and {Zhu}, Zhaohuan and {Birnstiel}, Tilman and {Bai}, Xue-Ning and {Benisty}, Myriam and {Hughes}, A. Meredith and {{\"O}berg}, Karin I. and {Ricci}, Luca},
        title = "{The Disk Substructures at High Angular Resolution Project (DSHARP). I. Motivation, Sample, Calibration, and Overview}",
      journal = {\apjl},
         year = 2018,
        month = dec,
       volume = {869},
       number = {2},
          eid = {L41},
        pages = {L41},
          doi = {10.3847/2041-8213/aaf741},
archivePrefix = {arXiv},
       eprint = {1812.04040},
 primaryClass = {astro-ph.SR},
       adsurl = {https://ui.adsabs.harvard.edu/abs/2018ApJ...869L..41A}
}

@ARTICLE{Birnstiel2018,
       author = {{Birnstiel}, Tilman and {Dullemond}, Cornelis P. and {Zhu}, Zhaohuan and {Andrews}, Sean M. and {Bai}, Xue-Ning and {Wilner}, David J. and {Carpenter}, John M. and {Huang}, Jane and {Isella}, Andrea and {Benisty}, Myriam and {P{\'e}rez}, Laura M. and {Zhang}, Shangjia},
        title = "{The Disk Substructures at High Angular Resolution Project (DSHARP). V. Interpreting ALMA Maps of Protoplanetary Disks in Terms of a Dust Model}",
      journal = {\apjl},
         year = 2018,
        month = dec,
       volume = {869},
       number = {2},
          eid = {L45},
        pages = {L45},
          doi = {10.3847/2041-8213/aaf743},
archivePrefix = {arXiv},
       eprint = {1812.04043},
 primaryClass = {astro-ph.SR},
       adsurl = {https://ui.adsabs.harvard.edu/abs/2018ApJ...869L..45B}
}

@ARTICLE{Carrasco-Gonzalez2016,
       author = {{Carrasco-Gonz{\'a}lez}, Carlos and {Henning}, Thomas and {Chandler}, Claire J. and {Linz}, Hendrik and {P{\'e}rez}, Laura and {Rodr{\'\i}guez}, Luis F. and {Galv{\'a}n-Madrid}, Roberto and {Anglada}, Guillem and {Birnstiel}, Til and {van Boekel}, Roy and {Flock}, Mario and {Klahr}, Hubert and {Macias}, Enrique and {Menten}, Karl and {Osorio}, Mayra and {Testi}, Leonardo and {Torrelles}, Jos{\'e} M. and {Zhu}, Zhaohuan},
        title = "{The VLA View of the HL Tau Disk: Disk Mass, Grain Evolution, and Early Planet Formation}",
      journal = {\apjl},
         year = 2016,
        month = apr,
       volume = {821},
       number = {1},
          eid = {L16},
        pages = {L16},
          doi = {10.3847/2041-8205/821/1/L16},
archivePrefix = {arXiv},
       eprint = {1603.03731},
 primaryClass = {astro-ph.SR},
       adsurl = {https://ui.adsabs.harvard.edu/abs/2016ApJ...821L..16C}
}

@ARTICLE{Carrasco-Gonzalez2019ApJ...883...71C,
       author = {{Carrasco-Gonz{\'a}lez}, Carlos and {Sierra}, Anibal and {Flock}, Mario and {Zhu}, Zhaohuan and {Henning}, Thomas and {Chandler}, Claire and {Galv{\'a}n-Madrid}, Roberto and {Mac{\'\i}as}, Enrique and {Anglada}, Guillem and {Linz}, Hendrik and {Osorio}, Mayra and {Rodr{\'\i}guez}, Luis F. and {Testi}, Leonardo and {Torrelles}, Jos{\'e} M. and {P{\'e}rez}, Laura and {Liu}, Yao},
        title = "{The Radial Distribution of Dust Particles in the HL Tau Disk from ALMA and VLA Observations}",
      journal = {\apj},
         year = 2019,
        month = sep,
       volume = {883},
       number = {1},
          eid = {71},
        pages = {71},
          doi = {10.3847/1538-4357/ab3d33},
archivePrefix = {arXiv},
       eprint = {1908.07140},
 primaryClass = {astro-ph.EP},
       adsurl = {https://ui.adsabs.harvard.edu/abs/2019ApJ...883...71C}
}

@ARTICLE{Chiang1997ApJ...490..368C,
       author = {{Chiang}, E.~I. and {Goldreich}, P.},
        title = "{Spectral Energy Distributions of T Tauri Stars with Passive Circumstellar Disks}",
      journal = {\apj},
         year = 1997,
        month = nov,
       volume = {490},
       number = {1},
        pages = {368-376},
          doi = {10.1086/304869},
archivePrefix = {arXiv},
       eprint = {astro-ph/9706042},
 primaryClass = {astro-ph},
       adsurl = {https://ui.adsabs.harvard.edu/abs/1997ApJ...490..368C}
}

@ARTICLE{Cho2007,
       author = {{Cho}, Jungyeon and {Lazarian}, A.},
        title = "{Grain Alignment and Polarized Emission from Magnetized T Tauri Disks}",
      journal = {\apj},
         year = 2007,
        month = nov,
       volume = {669},
       number = {2},
        pages = {1085-1097},
          doi = {10.1086/521805},
archivePrefix = {arXiv},
       eprint = {astro-ph/0611280},
 primaryClass = {astro-ph},
       adsurl = {https://ui.adsabs.harvard.edu/abs/2007ApJ...669.1085C}
}

@ARTICLE{Daniel1980,
       author = {{Daniel}, J.-Y.},
        title = "{Monte Carlo analysis of polarization by Mie scattering in circumstellar envelopes}",
      journal = {\aap},
         year = 1980,
        month = jul,
       volume = {87},
       number = {1-2},
        pages = {204-209},
       adsurl = {https://ui.adsabs.harvard.edu/abs/1980A&A....87..204D}
}

@software{Dominik2021ascl.soft04010D,
       author = {{Dominik}, Carsten and {Min}, Michiel and {Tazaki}, Ryo},
        title = "{OpTool: Command-line driven tool for creating complex dust opacities}",
 howpublished = {Astrophysics Source Code Library, record ascl:2104.010},
         year = 2021,
        month = apr,
          eid = {ascl:2104.010},
       adsurl = {https://ui.adsabs.harvard.edu/abs/2021ascl.soft04010D}
}

@INPROCEEDINGS{Drazkowska2023ASPC,
       author = {{Dr{\k{a}}{\.z}kowska}, J. and {Bitsch}, B. and {Lambrechts}, M. and {Mulders}, G.~D. and {Harsono}, D. and {Vazan}, A. and {Liu}, B. and {Ormel}, C.~W. and {Kretke}, K. and {Morbidelli}, A.},
        title = "{Planet Formation Theory in the Era of ALMA and Kepler: from Pebbles to Exoplanets}",
    booktitle = {Protostars and Planets VII},
         year = 2023,
       editor = {{Inutsuka}, S. and {Aikawa}, Y. and {Muto}, T. and {Tomida}, K. and {Tamura}, M.},
       series = {Astronomical Society of the Pacific Conference Series},
       volume = {534},
        month = jul,
        pages = {717},
          doi = {10.48550/arXiv.2203.09759},
archivePrefix = {arXiv},
       eprint = {2203.09759},
 primaryClass = {astro-ph.EP},
       adsurl = {https://ui.adsabs.harvard.edu/abs/2023ASPC..534..717D}
}

@software{Dullemond2012ascl.soft02015D,
       author = {{Dullemond}, C.~P. and {Juhasz}, A. and {Pohl}, A. and {Sereshti}, F. and {Shetty}, R. and {Peters}, T. and {Commercon}, B. and {Flock}, M.},
        title = "{RADMC-3D: A multi-purpose radiative transfer tool}",
 howpublished = {Astrophysics Source Code Library, record ascl:1202.015},
         year = 2012,
        month = feb,
          eid = {ascl:1202.015},
       adsurl = {https://ui.adsabs.harvard.edu/abs/2012ascl.soft02015D}
}

@ARTICLE{Escobar-Cerezo2017ApJ,
       author = {{Escobar-Cerezo}, J. and {Palmer}, C. and {Mu{\~n}oz}, O. and {Moreno}, F. and {Penttil{\"a}}, A. and {Muinonen}, K.},
        title = "{Scattering Properties of Large Irregular Cosmic Dust Particles at Visible Wavelengths}",
      journal = {\apj},
         year = 2017,
        month = mar,
       volume = {838},
       number = {1},
          eid = {74},
        pages = {74},
          doi = {10.3847/1538-4357/aa6303},
       adsurl = {https://ui.adsabs.harvard.edu/abs/2017ApJ...838...74E}
}

@ARTICLE{Haffert2019,
       author = {{Haffert}, S.~Y. and {Bohn}, A.~J. and {de Boer}, J. and {Snellen}, I.~A.~G. and {Brinchmann}, J. and {Girard}, J.~H. and {Keller}, C.~U. and {Bacon}, R.},
        title = "{Two accreting protoplanets around the young star PDS 70}",
      journal = {Nature Astronomy},
         year = 2019,
        month = jun,
       volume = {3},
        pages = {749-754},
          doi = {10.1038/s41550-019-0780-5},
archivePrefix = {arXiv},
       eprint = {1906.01486},
 primaryClass = {astro-ph.EP},
       adsurl = {https://ui.adsabs.harvard.edu/abs/2019NatAs...3..749H}
}

@ARTICLE{Harrison2019ApJ...877L...2H,
       author = {{Harrison}, Rachel E. and {Looney}, Leslie W. and {Stephens}, Ian W. and {Li}, Zhi-Yun and {Yang}, Haifeng and {Kataoka}, Akimasa and {Harris}, Robert J. and {Kwon}, Woojin and {Muto}, Takayuki and {Momose}, Munetake},
        title = "{Dust Polarization in Four Protoplanetary Disks at 3 mm: Further Evidence of Multiple Origins}",
      journal = {\apjl},
         year = 2019,
        month = may,
       volume = {877},
       number = {1},
          eid = {L2},
        pages = {L2},
          doi = {10.3847/2041-8213/ab1e46},
archivePrefix = {arXiv},
       eprint = {1905.06266},
 primaryClass = {astro-ph.SR},
       adsurl = {https://ui.adsabs.harvard.edu/abs/2019ApJ...877L...2H}
}

@ARTICLE{Harrison2024,
       author = {{Harrison}, Rachel E. and {Lin}, Zhe-Yu Daniel and {Looney}, Leslie W. and {Li}, Zhi-Yun and {Yang}, Haifeng and {Stephens}, Ian W. and {Fern{\'a}ndez-L{\'o}pez}, Manuel},
        title = "{Protoplanetary Disk Polarization at Multiple Wavelengths: Are Dust Populations Diverse?}",
      journal = {\apj},
         year = 2024,
        month = may,
       volume = {967},
       number = {1},
          eid = {40},
        pages = {40},
          doi = {10.3847/1538-4357/ad39ec},
archivePrefix = {arXiv},
       eprint = {2404.10217},
 primaryClass = {astro-ph.SR},
       adsurl = {https://ui.adsabs.harvard.edu/abs/2024ApJ...967...40H}
}

@BOOK{Hovenier2004ASSL,
       author = {{Hovenier}, Joop W. and {Van Der Mee}, Cornelis and {Domke}, Helmut},
        title = "{Transfer of polarized light in planetary atmospheres : basic concepts and practical methods}",
         year = 2004,
       volume = {318},
          doi = {10.1007/978-1-4020-2856-4},
        publisher = {Springer Dordrecht},
       adsurl = {https://ui.adsabs.harvard.edu/abs/2004ASSL..318.....H}
}

@ARTICLE{Hughes2009,
       author = {{Hughes}, A. Meredith and {Wilner}, David J. and {Cho}, Jungyeon and {Marrone}, Daniel P. and {Lazarian}, Alexandre and {Andrews}, Sean M. and {Rao}, Ramprasad},
        title = "{Stringent Limits on the Polarized Submillimeter Emission from Protoplanetary Disks}",
      journal = {\apj},
         year = 2009,
        month = oct,
       volume = {704},
       number = {2},
        pages = {1204-1217},
          doi = {10.1088/0004-637X/704/2/1204},
archivePrefix = {arXiv},
       eprint = {0909.1345},
 primaryClass = {astro-ph.SR},
       adsurl = {https://ui.adsabs.harvard.edu/abs/2009ApJ...704.1204H}
}

@ARTICLE{Hull2018ApJ...860...82H,
       author = {{Hull}, Charles L.~H. and {Yang}, Haifeng and {Li}, Zhi-Yun and {Kataoka}, Akimasa and {Stephens}, Ian W. and {Andrews}, Sean and {Bai}, Xuening and {Cleeves}, L. Ilsedore and {Hughes}, A. Meredith and {Looney}, Leslie and {P{\'e}rez}, Laura M. and {Wilner}, David},
        title = "{ALMA Observations of Polarization from Dust Scattering in the IM Lup Protoplanetary Disk}",
      journal = {\apj},
         year = 2018,
        month = jun,
       volume = {860},
       number = {1},
          eid = {82},
        pages = {82},
          doi = {10.3847/1538-4357/aabfeb},
archivePrefix = {arXiv},
       eprint = {1804.06269},
 primaryClass = {astro-ph.SR},
       adsurl = {https://ui.adsabs.harvard.edu/abs/2018ApJ...860...82H}
}

@ARTICLE{Kataoka2015,
       author = {{Kataoka}, Akimasa and {Muto}, Takayuki and {Momose}, Munetake and {Tsukagoshi}, Takashi and {Fukagawa}, Misato and {Shibai}, Hiroshi and {Hanawa}, Tomoyuki and {Murakawa}, Koji and {Dullemond}, Cornelis P.},
        title = "{Millimeter-wave Polarization of Protoplanetary Disks due to Dust Scattering}",
      journal = {\apj},
         year = 2015,
        month = aug,
       volume = {809},
       number = {1},
          eid = {78},
        pages = {78},
          doi = {10.1088/0004-637X/809/1/78},
archivePrefix = {arXiv},
       eprint = {1504.04812},
 primaryClass = {astro-ph.EP},
       adsurl = {https://ui.adsabs.harvard.edu/abs/2015ApJ...809...78K}
}

@ARTICLE{Kataoka2016a,
       author = {{Kataoka}, Akimasa and {Muto}, Takayuki and {Momose}, Munetake and {Tsukagoshi}, Takashi and {Dullemond}, Cornelis P.},
        title = "{Grain Size Constraints on HL Tau with Polarization Signature}",
      journal = {\apj},
         year = 2016,
        month = mar,
       volume = {820},
       number = {1},
          eid = {54},
        pages = {54},
          doi = {10.3847/0004-637X/820/1/54},
archivePrefix = {arXiv},
       eprint = {1507.08902},
 primaryClass = {astro-ph.EP},
       adsurl = {https://ui.adsabs.harvard.edu/abs/2016ApJ...820...54K}
}

@ARTICLE{Kataoka2016b,
       author = {{Kataoka}, Akimasa and {Tsukagoshi}, Takashi and {Momose}, Munetake and {Nagai}, Hiroshi and {Muto}, Takayuki and {Dullemond}, Cornelis P. and {Pohl}, Adriana and {Fukagawa}, Misato and {Shibai}, Hiroshi and {Hanawa}, Tomoyuki and {Murakawa}, Koji},
        title = "{Submillimeter Polarization Observation of the Protoplanetary Disk around HD 142527}",
      journal = {\apjl},
         year = 2016,
        month = nov,
       volume = {831},
       number = {2},
          eid = {L12},
        pages = {L12},
          doi = {10.3847/2041-8205/831/2/L12},
archivePrefix = {arXiv},
       eprint = {1610.06318},
 primaryClass = {astro-ph.EP},
       adsurl = {https://ui.adsabs.harvard.edu/abs/2016ApJ...831L..12K}
}

@ARTICLE{Kataoka2017,
       author = {{Kataoka}, Akimasa and {Tsukagoshi}, Takashi and {Pohl}, Adriana and {Muto}, Takayuki and {Nagai}, Hiroshi and {Stephens}, Ian W. and {Tomisaka}, Kohji and {Momose}, Munetake},
        title = "{The Evidence of Radio Polarization Induced by the Radiative Grain Alignment and Self-scattering of Dust Grains in a Protoplanetary Disk}",
      journal = {\apjl},
         year = 2017,
        month = jul,
       volume = {844},
       number = {1},
          eid = {L5},
        pages = {L5},
          doi = {10.3847/2041-8213/aa7e33},
archivePrefix = {arXiv},
       eprint = {1707.01612},
 primaryClass = {astro-ph.EP},
       adsurl = {https://ui.adsabs.harvard.edu/abs/2017ApJ...844L...5K}
}

@ARTICLE{Kataoka2019,
       author = {{Kataoka}, Akimasa and {Okuzumi}, Satoshi and {Tazaki}, Ryo},
        title = "{Millimeter-wave Polarization Due to Grain Alignment by the Gas Flow in Protoplanetary Disks}",
      journal = {\apjl},
         year = 2019,
        month = mar,
       volume = {874},
       number = {1},
          eid = {L6},
        pages = {L6},
          doi = {10.3847/2041-8213/ab0c9a},
archivePrefix = {arXiv},
       eprint = {1903.03529},
 primaryClass = {astro-ph.EP},
       adsurl = {https://ui.adsabs.harvard.edu/abs/2019ApJ...874L...6K}
}

@ARTICLE{Keppler2018,
       author = {{Keppler}, M. and {Benisty}, M. and {M{\"u}ller}, A. and {Henning}, Th. and {van Boekel}, R. and {Cantalloube}, F. and {Ginski}, C. and {van Holstein}, R.~G. and {Maire}, A. -L. and {Pohl}, A. and {Samland}, M. and {Avenhaus}, H. and {Baudino}, J. -L. and {Boccaletti}, A. and {de Boer}, J. and {Bonnefoy}, M. and {Chauvin}, G. and {Desidera}, S. and {Langlois}, M. and {Lazzoni}, C. and {Marleau}, G. -D. and {Mordasini}, C. and {Pawellek}, N. and {Stolker}, T. and {Vigan}, A. and {Zurlo}, A. and {Birnstiel}, T. and {Brandner}, W. and {Feldt}, M. and {Flock}, M. and {Girard}, J. and {Gratton}, R. and {Hagelberg}, J. and {Isella}, A. and {Janson}, M. and {Juhasz}, A. and {Kemmer}, J. and {Kral}, Q. and {Lagrange}, A. -M. and {Launhardt}, R. and {Matter}, A. and {M{\'e}nard}, F. and {Milli}, J. and {Molli{\`e}re}, P. and {Olofsson}, J. and {P{\'e}rez}, L. and {Pinilla}, P. and {Pinte}, C. and {Quanz}, S.~P. and {Schmidt}, T. and {Udry}, S. and {Wahhaj}, Z. and {Williams}, J.~P. and {Buenzli}, E. and {Cudel}, M. and {Dominik}, C. and {Galicher}, R. and {Kasper}, M. and {Lannier}, J. and {Mesa}, D. and {Mouillet}, D. and {Peretti}, S. and {Perrot}, C. and {Salter}, G. and {Sissa}, E. and {Wildi}, F. and {Abe}, L. and {Antichi}, J. and {Augereau}, J. -C. and {Baruffolo}, A. and {Baudoz}, P. and {Bazzon}, A. and {Beuzit}, J. -L. and {Blanchard}, P. and {Brems}, S.~S. and {Buey}, T. and {De Caprio}, V. and {Carbillet}, M. and {Carle}, M. and {Cascone}, E. and {Cheetham}, A. and {Claudi}, R. and {Costille}, A. and {Delboulb{\'e}}, A. and {Dohlen}, K. and {Fantinel}, D. and {Feautrier}, P. and {Fusco}, T. and {Giro}, E. and {Gluck}, L. and {Gry}, C. and {Hubin}, N. and {Hugot}, E. and {Jaquet}, M. and {Le Mignant}, D. and {Llored}, M. and {Madec}, F. and {Magnard}, Y. and {Martinez}, P. and {Maurel}, D. and {Meyer}, M. and {M{\"o}ller-Nilsson}, O. and {Moulin}, T. and {Mugnier}, L. and {Orign{\'e}}, A. and {Pavlov}, A. and {Perret}, D. and {Petit}, C. and {Pragt}, J. and {Puget}, P. and {Rabou}, P. and {Ramos}, J. and {Rigal}, F. and {Rochat}, S. and {Roelfsema}, R. and {Rousset}, G. and {Roux}, A. and {Salasnich}, B. and {Sauvage}, J. -F. and {Sevin}, A. and {Soenke}, C. and {Stadler}, E. and {Suarez}, M. and {Turatto}, M. and {Weber}, L.},
        title = "{Discovery of a planetary-mass companion within the gap of the transition disk around PDS 70}",
      journal = {\aap},
         year = 2018,
        month = sep,
       volume = {617},
          eid = {A44},
        pages = {A44},
          doi = {10.1051/0004-6361/201832957},
archivePrefix = {arXiv},
       eprint = {1806.11568},
 primaryClass = {astro-ph.EP},
       adsurl = {https://ui.adsabs.harvard.edu/abs/2018A&A...617A..44K}
}

@ARTICLE{Kirchschlager2020,
       author = {{Kirchschlager}, Florian and {Bertrang}, Gesa H. -M.},
        title = "{Self-scattering of non-spherical dust grains. The limitations of perfect compact spheres}",
      journal = {\aap},
         year = 2020,
        month = jun,
       volume = {638},
          eid = {A116},
        pages = {A116},
          doi = {10.1051/0004-6361/202037943},
archivePrefix = {arXiv},
       eprint = {2004.13742},
 primaryClass = {astro-ph.SR},
       adsurl = {https://ui.adsabs.harvard.edu/abs/2020A&A...638A.116K}
}

@ARTICLE{Kirchschlager2019,
       author = {{Kirchschlager}, Florian and {Bertrang}, Gesa H. -M. and {Flock}, Mario},
        title = "{Intrinsic polarization of elongated porous dust grains}",
      journal = {\mnras},
         year = 2019,
        month = sep,
       volume = {488},
       number = {1},
        pages = {1211-1219},
          doi = {10.1093/mnras/stz1763},
archivePrefix = {arXiv},
       eprint = {1906.10699},
 primaryClass = {astro-ph.EP},
       adsurl = {https://ui.adsabs.harvard.edu/abs/2019MNRAS.488.1211K}
}

@ARTICLE{Lin2023,
       author = {{Lin}, Zhe-Yu Daniel and {Li}, Zhi-Yun and {Yang}, Haifeng and {Mu{\~n}oz}, Olga and {Looney}, Leslie and {Stephens}, Ian and {Hull}, Charles L.~H. and {Fern{\'a}ndez-L{\'o}pez}, Manuel and {Harrison}, Rachel},
        title = "{(Sub)millimetre dust polarization of protoplanetary discs from scattering by large millimetre-sized irregular grains}",
      journal = {\mnras},
         year = 2023,
        month = mar,
       volume = {520},
       number = {1},
        pages = {1210-1223},
          doi = {10.1093/mnras/stad173},
archivePrefix = {arXiv},
       eprint = {2206.12357},
 primaryClass = {astro-ph.EP},
       adsurl = {https://ui.adsabs.harvard.edu/abs/2023MNRAS.520.1210L}
}

@ARTICLE{Lin2024,
       author = {{Lin}, Zhe-Yu Daniel and {Li}, Zhi-Yun and {Stephens}, Ian W. and {Fern{\'a}ndez-L{\'o}pez}, Manuel and {Carrasco-Gonz{\'a}lez}, Carlos and {Chandler}, Claire J. and {Pasetto}, Alice and {Looney}, Leslie W. and {Yang}, Haifeng and {Harrison}, Rachel E. and {Sadavoy}, Sarah I. and {Henning}, Thomas and {Hughes}, A. Meredith and {Kataoka}, Akimasa and {Kwon}, Woojin and {Muto}, Takayuki and {Segura-Cox}, Dominique},
        title = "{Panchromatic (Sub)millimeter polarization observations of HL Tau unveil aligned scattering grains}",
      journal = {\mnras},
         year = 2024,
        month = feb,
       volume = {528},
       number = {1},
        pages = {843-862},
          doi = {10.1093/mnras/stae040},
archivePrefix = {arXiv},
       eprint = {2309.10055},
 primaryClass = {astro-ph.EP},
       adsurl = {https://ui.adsabs.harvard.edu/abs/2024MNRAS.528..843L}
}

@ARTICLE{Lin2024aligned,
       author = {{Lin}, Zhe-Yu Daniel and {Li}, Zhi-Yun and {Yang}, Haifeng and {Looney}, Leslie W. and {Stephens}, Ian W. and {Fern{\'a}ndez-L{\'o}pez}, Manuel and {Harrison}, Rachel E.},
        title = "{Badminton birdie-like aerodynamic alignment of drifting dust grains by subsonic gaseous flows in protoplanetary discs}",
      journal = {\mnras},
         year = 2024,
        month = nov,
       volume = {534},
       number = {4},
        pages = {3713-3733},
          doi = {10.1093/mnras/stae2248},
archivePrefix = {arXiv},
       eprint = {2407.10025},
 primaryClass = {astro-ph.EP},
       adsurl = {https://ui.adsabs.harvard.edu/abs/2024MNRAS.534.3713L}
}

@ARTICLE{Liu2024,
       author = {{Liu}, Yao and {Roussel}, H{\'e}l{\`e}ne and {Linz}, Hendrik and {Fang}, Min and {Wolf}, Sebastian and {Kirchschlager}, Florian and {Henning}, Thomas and {Yang}, Haifeng and {Du}, Fujun and {Flock}, Mario and {Wang}, Hongchi},
        title = "{Dust mass in protoplanetary disks with porous dust opacities}",
      journal = {\aap},
         year = 2024,
        month = dec,
       volume = {692},
          eid = {A148},
        pages = {A148},
          doi = {10.1051/0004-6361/202451981},
archivePrefix = {arXiv},
       eprint = {2411.00277},
 primaryClass = {astro-ph.EP},
       adsurl = {https://ui.adsabs.harvard.edu/abs/2024A&A...692A.148L}
}

@ARTICLE{Long2018,
       author = {{Long}, Feng and {Pinilla}, Paola and {Herczeg}, Gregory J. and {Harsono}, Daniel and {Dipierro}, Giovanni and {Pascucci}, Ilaria and {Hendler}, Nathan and {Tazzari}, Marco and {Ragusa}, Enrico and {Salyk}, Colette and {Edwards}, Suzan and {Lodato}, Giuseppe and {van de Plas}, Gerrit and {Johnstone}, Doug and {Liu}, Yao and {Boehler}, Yann and {Cabrit}, Sylvie and {Manara}, Carlo F. and {Menard}, Francois and {Mulders}, Gijs D. and {Nisini}, Brunella and {Fischer}, William J. and {Rigliaco}, Elisabetta and {Banzatti}, Andrea and {Avenhaus}, Henning and {Gully-Santiago}, Michael},
        title = "{Gaps and Rings in an ALMA Survey of Disks in the Taurus Star-forming Region}",
      journal = {\apj},
         year = 2018,
        month = dec,
       volume = {869},
       number = {1},
          eid = {17},
        pages = {17},
          doi = {10.3847/1538-4357/aae8e1},
archivePrefix = {arXiv},
       eprint = {1810.06044},
 primaryClass = {astro-ph.SR},
       adsurl = {https://ui.adsabs.harvard.edu/abs/2018ApJ...869...17L}
}

@ARTICLE{Lynden-Bell1974MNRAS.168..603L,
       author = {{Lynden-Bell}, D. and {Pringle}, J.~E.},
        title = "{The evolution of viscous discs and the origin of the nebular variables.}",
      journal = {\mnras},
         year = 1974,
        month = sep,
       volume = {168},
        pages = {603-637},
          doi = {10.1093/mnras/168.3.603},
       adsurl = {https://ui.adsabs.harvard.edu/abs/1974MNRAS.168..603L}
}

@ARTICLE{Macias2021A&A...648A..33M,
       author = {{Mac{\'\i}as}, E. and {Guerra-Alvarado}, O. and {Carrasco-Gonz{\'a}lez}, C. and {Ribas}, {\'A}. and {Espaillat}, C.~C. and {Huang}, J. and {Andrews}, S.~M.},
        title = "{Characterizing the dust content of disk substructures in TW Hydrae}",
      journal = {\aap},
         year = 2021,
        month = apr,
       volume = {648},
          eid = {A33},
        pages = {A33},
          doi = {10.1051/0004-6361/202039812},
archivePrefix = {arXiv},
       eprint = {2102.04648},
 primaryClass = {astro-ph.EP},
       adsurl = {https://ui.adsabs.harvard.edu/abs/2021A&A...648A..33M}
}

@ARTICLE{Martikainen2025MNRAS.537.1489M,
       author = {{Martikainen}, Julia and {Mu{\~n}oz}, Olga and {G{\'o}mez Mart{\'\i}n}, Juan Carlos and {Passas Varo}, Mar{\'\i}a and {Jardiel}, Teresa and {Peiteado}, Marco and {Willame}, Yannick and {Neary}, Lori and {Becker}, Tim and {Wurm}, Gerhard},
        title = "{Database of Martian dust optical properties in the UV-vis-NIR}",
      journal = {\mnras},
         year = 2025,
        month = feb,
       volume = {537},
       number = {2},
        pages = {1489-1503},
          doi = {10.1093/mnras/staf108},
       adsurl = {https://ui.adsabs.harvard.edu/abs/2025MNRAS.537.1489M}
}

@book{Mishchenko2000,
  editor    = {Michael I. Mishchenko and Joop W. Hovenier and Larry D. Travis},
  title     = {Light Scattering by Nonspherical Particles: Theory, Measurements, and Applications},
  publisher = {Academic Press},
  year      = {2000},
  address   = {San Diego},
  isbn      = {978-0-12-498660-2}
}

@article{MUINONEN20091628,
title = {Light scattering by Gaussian particles with internal inclusions and roughened surfaces using ray optics},
journal = {Journal of Quantitative Spectroscopy and Radiative Transfer},
volume = {110},
number = {14},
pages = {1628-1639},
year = {2009},
note = {XI Conference on Electromagnetic and Light Scattering by Non-Spherical Particles: 2008},
issn = {0022-4073},
doi = {https://doi.org/10.1016/j.jqsrt.2009.03.012},
url = {https://www.sciencedirect.com/science/article/pii/S0022407309000983},
author = {Karri Muinonen and Timo Nousiainen and Hannakaisa Lindqvist and Olga Muñoz and Gorden Videen}
}

@ARTICLE{Muller2018,
       author = {{M{\"u}ller}, A. and {Keppler}, M. and {Henning}, Th. and {Samland}, M. and {Chauvin}, G. and {Beust}, H. and {Maire}, A. -L. and {Molaverdikhani}, K. and {van Boekel}, R. and {Benisty}, M. and {Boccaletti}, A. and {Bonnefoy}, M. and {Cantalloube}, F. and {Charnay}, B. and {Baudino}, J. -L. and {Gennaro}, M. and {Long}, Z.~C. and {Cheetham}, A. and {Desidera}, S. and {Feldt}, M. and {Fusco}, T. and {Girard}, J. and {Gratton}, R. and {Hagelberg}, J. and {Janson}, M. and {Lagrange}, A. -M. and {Langlois}, M. and {Lazzoni}, C. and {Ligi}, R. and {M{\'e}nard}, F. and {Mesa}, D. and {Meyer}, M. and {Molli{\`e}re}, P. and {Mordasini}, C. and {Moulin}, T. and {Pavlov}, A. and {Pawellek}, N. and {Quanz}, S.~P. and {Ramos}, J. and {Rouan}, D. and {Sissa}, E. and {Stadler}, E. and {Vigan}, A. and {Wahhaj}, Z. and {Weber}, L. and {Zurlo}, A.},
        title = "{Orbital and atmospheric characterization of the planet within the gap of the PDS 70 transition disk}",
      journal = {\aap},
         year = 2018,
        month = sep,
       volume = {617},
          eid = {L2},
        pages = {L2},
          doi = {10.1051/0004-6361/201833584},
archivePrefix = {arXiv},
       eprint = {1806.11567},
 primaryClass = {astro-ph.EP},
       adsurl = {https://ui.adsabs.harvard.edu/abs/2018A&A...617L...2M}
}

@ARTICLE{Munoz2021ApJS..256...17M,
       author = {{Mu{\~n}oz}, O. and {Frattin}, E. and {Jardiel}, T. and {G{\'o}mez-Mart{\'\i}n}, J.~C. and {Moreno}, F. and {Ramos}, J.~L. and {Guirado}, D. and {Peiteado}, M. and {Caballero}, A.~C. and {Milli}, J. and {M{\'e}nard}, F.},
        title = "{Retrieving Dust Grain Sizes from Photopolarimetry: An Experimental Approach}",
      journal = {\apjs},
         year = 2021,
        month = sep,
       volume = {256},
       number = {1},
          eid = {17},
        pages = {17},
          doi = {10.3847/1538-4365/ac0efa},
archivePrefix = {arXiv},
       eprint = {2109.05764},
 primaryClass = {astro-ph.EP},
       adsurl = {https://ui.adsabs.harvard.edu/abs/2021ApJS..256...17M}
}

@ARTICLE{Ohashi2018,
       author = {{Ohashi}, Satoshi and {Kataoka}, Akimasa and {Nagai}, Hiroshi and {Momose}, Munetake and {Muto}, Takayuki and {Hanawa}, Tomoyuki and {Fukagawa}, Misato and {Tsukagoshi}, Takashi and {Murakawa}, Kohji and {Shibai}, Hiroshi},
        title = "{Two Different Grain Size Distributions within the Protoplanetary Disk around HD 142527 Revealed by ALMA Polarization Observation}",
      journal = {\apj},
         year = 2018,
        month = sep,
       volume = {864},
       number = {1},
          eid = {81},
        pages = {81},
          doi = {10.3847/1538-4357/aad632},
archivePrefix = {arXiv},
       eprint = {1807.10776},
 primaryClass = {astro-ph.EP},
       adsurl = {https://ui.adsabs.harvard.edu/abs/2018ApJ...864...81O}
}

@ARTICLE{Perez2012,
       author = {{P{\'e}rez}, Laura M. and {Carpenter}, John M. and {Chandler}, Claire J. and {Isella}, Andrea and {Andrews}, Sean M. and {Ricci}, Luca and {Calvet}, Nuria and {Corder}, Stuartt A. and {Deller}, Adam T. and {Dullemond}, Cornelis P. and {Greaves}, Jane S. and {Harris}, Robert J. and {Henning}, Thomas and {Kwon}, Woojin and {Lazio}, Joseph and {Linz}, Hendrik and {Mundy}, Lee G. and {Sargent}, Anneila I. and {Storm}, Shaye and {Testi}, Leonardo and {Wilner}, David J.},
        title = "{Constraints on the Radial Variation of Grain Growth in the AS 209 Circumstellar Disk}",
      journal = {\apjl},
         year = 2012,
        month = nov,
       volume = {760},
       number = {1},
          eid = {L17},
        pages = {L17},
          doi = {10.1088/2041-8205/760/1/L17},
archivePrefix = {arXiv},
       eprint = {1210.5252},
 primaryClass = {astro-ph.SR},
       adsurl = {https://ui.adsabs.harvard.edu/abs/2012ApJ...760L..17P}
}

@ARTICLE{Potapov2025,
       author = {{Potapov}, Alexey and {McCoustra}, Martin R.~S. and {Tazaki}, Ryo and {Bergin}, Edwin A. and {Bromley}, Stefan T. and {Garrod}, Robin T. and {Rimola}, Albert},
        title = "{Is cosmic dust porous?}",
      journal = {\aapr},
         year = 2025,
        month = oct,
       volume = {33},
       number = {1},
          eid = {6},
        pages = {6},
          doi = {10.1007/s00159-025-00164-5},
archivePrefix = {arXiv},
       eprint = {2509.10292},
 primaryClass = {astro-ph.GA},
       adsurl = {https://ui.adsabs.harvard.edu/abs/2025A&ARv..33....6P}
}

@ARTICLE{ReyesAmador2024,
       author = {{Reyes-Amador}, Omar Ulises and {Fritz}, Jacopo and {Gonz{\'a}lez-Mart{\'\i}n}, Omaira and {Srinivasan}, Sundar and {Baes}, Maarten and {Lopez-Rodriguez}, Enrique and {Osorio-Clavijo}, Natalia and {Victoria-Ceballos}, Cesar Iv{\'a}n and {Stalevski}, Marko and {Ramos Almeida}, C.},
        title = "{Towards an observationally motivated AGN dusty torus model - I. Dust chemical composition from the modelling of Spitzer spectra}",
      journal = {\mnras},
         year = 2024,
        month = jun,
       volume = {531},
       number = {1},
        pages = {1841-1855},
          doi = {10.1093/mnras/stae1281},
archivePrefix = {arXiv},
       eprint = {2405.08960},
 primaryClass = {astro-ph.GA},
       adsurl = {https://ui.adsabs.harvard.edu/abs/2024MNRAS.531.1841R}
}

@ARTICLE{Saito2021JAtS...78.2089S,
       author = {{Saito}, Masanori and {Yang}, Ping and {Ding}, Jiachen and {Liu}, Xu},
        title = "{A Comprehensive Database of the Optical Properties of Irregular Aerosol Particles for Radiative Transfer Simulations}",
      journal = {Journal of the Atmospheric Sciences},
         year = 2021,
        month = jul,
       volume = {78},
       number = {7},
        pages = {2089-2111},
          doi = {10.1175/JAS-D-20-0338.1},
       adsurl = {https://ui.adsabs.harvard.edu/abs/2021JAtS...78.2089S}
}

@dataset{masanori_saito_2021_4711247,
  author       = {Masanori Saito},
  title        = {TAMUdust2020 Database},
  month        = apr,
  year         = 2021,
  publisher    = {Zenodo},
  version      = {1.0.0},
  doi          = {10.5281/zenodo.4711247},
  url          = {https://doi.org/10.5281/zenodo.4711247},
}

@ARTICLE{Sierra2020ApJ...892..136S,
       author = {{Sierra}, Anibal and {Lizano}, Susana},
        title = "{Effects of Scattering, Temperature Gradients, and Settling on the Derived Dust Properties of Observed Protoplanetary Disks}",
      journal = {\apj},
         year = 2020,
        month = apr,
       volume = {892},
       number = {2},
          eid = {136},
        pages = {136},
          doi = {10.3847/1538-4357/ab7d32},
archivePrefix = {arXiv},
       eprint = {2003.02982},
 primaryClass = {astro-ph.EP},
       adsurl = {https://ui.adsabs.harvard.edu/abs/2020ApJ...892..136S}
}

@ARTICLE{Stephens2017,
       author = {{Stephens}, Ian W. and {Yang}, Haifeng and {Li}, Zhi-Yun and {Looney}, Leslie W. and {Kataoka}, Akimasa and {Kwon}, Woojin and {Fern{\'a}ndez-L{\'o}pez}, Manuel and {Hull}, Charles L.~H. and {Hughes}, Meredith and {Segura-Cox}, Dominique and {Mundy}, Lee and {Crutcher}, Richard and {Rao}, Ramprasad},
        title = "{ALMA Reveals Transition of Polarization Pattern with Wavelength in HL Tau{\textquoteright}s Disk}",
      journal = {\apj},
         year = 2017,
        month = dec,
       volume = {851},
       number = {1},
          eid = {55},
        pages = {55},
          doi = {10.3847/1538-4357/aa998b},
archivePrefix = {arXiv},
       eprint = {1710.04670},
 primaryClass = {astro-ph.SR},
       adsurl = {https://ui.adsabs.harvard.edu/abs/2017ApJ...851...55S}
}

@ARTICLE{Stephens2023,
       author = {{Stephens}, Ian W. and {Lin}, Zhe-Yu Daniel and {Fern{\'a}ndez-L{\'o}pez}, Manuel and {Li}, Zhi-Yun and {Looney}, Leslie W. and {Yang}, Haifeng and {Harrison}, Rachel and {Kataoka}, Akimasa and {Carrasco-Gonzalez}, Carlos and {Okuzumi}, Satoshi and {Tazaki}, Ryo},
        title = "{Aligned grains and scattered light found in gaps of planet-forming disk}",
      journal = {\nat},
         year = 2023,
        month = nov,
       volume = {623},
       number = {7988},
        pages = {705-708},
          doi = {10.1038/s41586-023-06648-7},
archivePrefix = {arXiv},
       eprint = {2311.08452},
 primaryClass = {astro-ph.GA},
       adsurl = {https://ui.adsabs.harvard.edu/abs/2023Natur.623..705S}
}

@ARTICLE{Tazaki2016,
       author = {{Tazaki}, Ryo and {Tanaka}, Hidekazu and {Okuzumi}, Satoshi and {Kataoka}, Akimasa and {Nomura}, Hideko},
        title = "{Light Scattering by Fractal Dust Aggregates. I. Angular Dependence of Scattering}",
      journal = {\apj},
         year = 2016,
        month = jun,
       volume = {823},
       number = {2},
          eid = {70},
        pages = {70},
          doi = {10.3847/0004-637X/823/2/70},
archivePrefix = {arXiv},
       eprint = {1603.07492},
 primaryClass = {astro-ph.EP},
       adsurl = {https://ui.adsabs.harvard.edu/abs/2016ApJ...823...70T}
}

@ARTICLE{Tazaki2017,
       author = {{Tazaki}, Ryo and {Lazarian}, Alexandre and {Nomura}, Hideko},
        title = "{Radiative Grain Alignment In Protoplanetary Disks: Implications for Polarimetric Observations}",
      journal = {\apj},
         year = 2017,
        month = apr,
       volume = {839},
       number = {1},
          eid = {56},
        pages = {56},
          doi = {10.3847/1538-4357/839/1/56},
archivePrefix = {arXiv},
       eprint = {1701.02063},
 primaryClass = {astro-ph.EP},
       adsurl = {https://ui.adsabs.harvard.edu/abs/2017ApJ...839...56T}
}

@ARTICLE{Tazaki2019,
       author = {{Tazaki}, Ryo and {Tanaka}, Hidekazu and {Kataoka}, Akimasa and {Okuzumi}, Satoshi and {Muto}, Takayuki},
        title = "{Unveiling Dust Aggregate Structure in Protoplanetary Disks by Millimeter-wave Scattering Polarization}",
      journal = {\apj},
         year = 2019,
        month = nov,
       volume = {885},
       number = {1},
          eid = {52},
        pages = {52},
          doi = {10.3847/1538-4357/ab45f0},
archivePrefix = {arXiv},
       eprint = {1907.00189},
 primaryClass = {astro-ph.EP},
       adsurl = {https://ui.adsabs.harvard.edu/abs/2019ApJ...885...52T}
}

@ARTICLE{Yang2016a,
       author = {{Yang}, Haifeng and {Li}, Zhi-Yun and {Looney}, Leslie and {Stephens}, Ian},
        title = "{Inclination-induced polarization of scattered millimetre radiation from protoplanetary discs: the case of HL Tau}",
      journal = {\mnras},
         year = 2016,
        month = mar,
       volume = {456},
       number = {3},
        pages = {2794-2805},
          doi = {10.1093/mnras/stv2633},
archivePrefix = {arXiv},
       eprint = {1507.08353},
 primaryClass = {astro-ph.SR},
       adsurl = {https://ui.adsabs.harvard.edu/abs/2016MNRAS.456.2794Y}
}

@ARTICLE{Yang2016b,
       author = {{Yang}, Haifeng and {Li}, Zhi-Yun and {Looney}, Leslie W. and {Cox}, Erin G. and {Tobin}, John and {Stephens}, Ian W. and {Segura-Cox}, Dominque M. and {Harris}, Robert J.},
        title = "{Disc polarization from both emission and scattering of magnetically aligned grains: the case of NGC 1333 IRAS 4A1}",
      journal = {\mnras},
         year = 2016,
        month = aug,
       volume = {460},
       number = {4},
        pages = {4109-4121},
          doi = {10.1093/mnras/stw1253},
archivePrefix = {arXiv},
       eprint = {1602.08196},
 primaryClass = {astro-ph.GA},
       adsurl = {https://ui.adsabs.harvard.edu/abs/2016MNRAS.460.4109Y}
}

@ARTICLE{Yang2019,
       author = {{Yang}, Haifeng and {Li}, Zhi-Yun and {Stephens}, Ian W. and {Kataoka}, Akimasa and {Looney}, Leslie},
        title = "{Does HL Tau disc polarization in ALMA band 3 come from radiatively aligned grains?}",
      journal = {\mnras},
         year = 2019,
        month = feb,
       volume = {483},
       number = {2},
        pages = {2371-2381},
          doi = {10.1093/mnras/sty3263},
archivePrefix = {arXiv},
       eprint = {1811.11897},
 primaryClass = {astro-ph.SR},
       adsurl = {https://ui.adsabs.harvard.edu/abs/2019MNRAS.483.2371Y}
}

@ARTICLE{Zhang2023,
       author = {{Zhang}, Shangjia and {Zhu}, Zhaohuan and {Ueda}, Takahiro and {Kataoka}, Akimasa and {Sierra}, Anibal and {Carrasco-Gonz{\'a}lez}, Carlos and {Mac{\'\i}as}, Enrique},
        title = "{Porous Dust Particles in Protoplanetary Disks: Application to the HL Tau Disk}",
      journal = {\apj},
         year = 2023,
        month = aug,
       volume = {953},
       number = {1},
          eid = {96},
        pages = {96},
          doi = {10.3847/1538-4357/acdb4e},
archivePrefix = {arXiv},
       eprint = {2306.00158},
 primaryClass = {astro-ph.EP},
       adsurl = {https://ui.adsabs.harvard.edu/abs/2023ApJ...953...96Z}
}

@article{astropy:2013,
Adsurl = {http://adsabs.harvard.edu/abs/2013A%26A...558A..33A},
Archiveprefix = {arXiv},
Author = {{Astropy Collaboration} and {Robitaille}, T.~P. and {Tollerud}, E.~J. and {Greenfield}, P. and {Droettboom}, M. and {Bray}, E. and {Aldcroft}, T. and {Davis}, M. and {Ginsburg}, A. and {Price-Whelan}, A.~M. and {Kerzendorf}, W.~E. and {Conley}, A. and {Crighton}, N. and {Barbary}, K. and {Muna}, D. and {Ferguson}, H. and {Grollier}, F. and {Parikh}, M.~M. and {Nair}, P.~H. and {Unther}, H.~M. and {Deil}, C. and {Woillez}, J. and {Conseil}, S. and {Kramer}, R. and {Turner}, J.~E.~H. and {Singer}, L. and {Fox}, R. and {Weaver}, B.~A. and {Zabalza}, V. and {Edwards}, Z.~I. and {Azalee Bostroem}, K. and {Burke}, D.~J. and {Casey}, A.~R. and {Crawford}, S.~M. and {Dencheva}, N. and {Ely}, J. and {Jenness}, T. and {Labrie}, K. and {Lim}, P.~L. and {Pierfederici}, F. and {Pontzen}, A. and {Ptak}, A. and {Refsdal}, B. and {Servillat}, M. and {Streicher}, O.},
Doi = {10.1051/0004-6361/201322068},
Eid = {A33},
Eprint = {1307.6212},
Journal = {\aap},
Month = oct,
Pages = {A33},
Primaryclass = {astro-ph.IM},
Title = {{Astropy: A community Python package for astronomy}},
Volume = 558,
Year = 2013}
\bibliographystyle{aasjournalv7}

\end{document}